\documentclass[12pt]{iopart}
\usepackage{amssymb}
\usepackage{color}
\usepackage{graphicx}

\def\gr{$\gamma$-ray} 

\begin{document}

\title[UHECR from polar caps in BH magnetospheres]{Ultra-High Energy Cosmic Ray production in the polar cap
    regions of black hole magnetospheres}

  \author{A.Yu.~Neronov$^{1,2}$, D.V.~Semikoz$^{3,4}$, I.I.~Tkachev$^{4}$} 
  \address{$^1$ISDC, CH. d'Ecogia 16, Verosox, 1290, Switzerland}
  \address{$^2$Geneva Observatory, Ch. des Maillettes, Versoix
    1290, Switzerland} \address{$^3$APC, 10, rue Alice Domon et
    Leonie Duquet, F--75205 Paris Cedex 13, France}
\address{$^4$Institute for Nuclear Researches of
    the Russian Academy of Sciences, 60th October Anniversary
    Prosp. 7a, 117312, Moscow, Russia}

\ead{Andrii.Neronov@unige.ch}
\begin{abstract}
  We develop a model of ultra-high energy cosmic ray (UHECR)
  production via acceleration in a rotation-induced electric field in
  vacuum gaps in the magnetospheres of supermassive black holes
  (BH). We show that if the poloidal magnetic field near the BH
  horizon is misaligned with the BH rotation axis, charged particles,
  which initially spiral into the BH hole along the equatorial plane,
  penetrate into the regions above the BH "polar caps" and are ejected
  with high energies to infinity.  We show that in such a model
  acceleration of protons near a BH of typical mass $3\times 10^8$
  solar masses is possible only if the magnetic field is almost
  aligned with the BH rotation axis. We find that the power of
  anisotropic electromagnetic emission from an UHECR source near a
  supermassive BH should be at least 10-100 times larger then UHECR power
  of the source.  This implies that if the number of UHECR sources
  within the 100~Mpc sphere is $\sim 100$, the power of
  electromagnetic emission which accompanies proton acceleration in
  each source, $10^{42-43}$~erg/s, is comparable to the typical
  luminosities of active galactic nuclei (AGN) in the local
  Universe. We also explore the acceleration of heavy nuclei, for
  which the constraints on the electromagnetic luminosity and on the
  alignment of magnetic field in the gap are relaxed.
\end{abstract}

\maketitle

\section{Introduction}

The existence of the Greisen-Zatsepin-Kuzmin (GZK) cutoff
\cite{greisen,zk} in the spectrum of UHECR, found by the HiRes
experiment~\cite{HiRes_GZK} and confirmed recently by Pierre Auger
Observatory~\cite{spec_AUGER}, points to the astrophysical origin of
the primary cosmic ray particles.  In this case most of the cosmic
rays with energies above the cut-off energy $E\simeq 10^{20}$ eV
should come from nearby sources located at the distance $D<100$
Mpc. At the energies $\sim 10^{20}$~eV the trajectories of the cosmic
ray protons are not significantly deflected by magnetic fields, so
that the UHECR should point directly to their sources. However, the
statistics of UHECR events was not enough, up to recently, to identify
the populaiton of astornomical sources, responsible for the UHECR
production.

The first attempt to identify the CR sources via a direct back-tracing
of UHECR events back to their sources was reported by the Peirre Auger
Observatory (PAO) collaboration \cite{auger_science,auger_app}, which
found an evidence for the correlation between the arrival directions
of the UHECR and directions toward the nearby AGN. The correlation
signal was found at a rather small angles with a large number of
different AGN contributing to it, and therefore it was interpreted as
caused by proton primaries.  However, the data presented in
\cite{auger_science,auger_app} are not really consistent with such AGN
hypothesis \cite{Gorbunov:2007ja} and correlations with AGN are absent
in the more recent HiRes data \cite{hires}. Moreover, a different
interpretation of the Auger correlation signal is possible
\cite{Gorbunov:2007ja,Wibig:2007pf,Fargion:2008sp,Gorbunov:2008ef},
with just a few sources (or even a single source such as Cen A)
contributing in the relevant part of the local Universe and with heavier
nuclei making substantial fraction of the flux. There is
yet another difficulty with the AGN interpretation of the Auger
signal. Local ANG are weak, hosting black holes of small mass in their
hearts, and without strong jets.  Acceleration of protons to highest
energies in such AGN is not expected, for a discussion of this issue
with respect to the Auger signal see
Refs. \cite{Moskalenko:2008iz,Ptitsyna:2008zs,Gureev:2008bj}.

It is clear that an increase of the event statistics is
will lead to further improvement of our knowledge of the astronomical
sources of UHECR.  It is, however, also clear that even with the
increase of the statistics (this requires accumulation of the signal
over many years) the sensitivity of the UHECR experiments would not be
enough to pin-point the individual UHECR sources. Even if the class of
the objects which produce the observed UHECR flux would be determined
(say, a certain type of AGN), only very limited information about the
location of the particle acceleration sites and physical conditions in
these sources would be extractable from the UHECR data alone.

Assuming that at least some of AGN are potential sources \cite{berezinsky},
different "candidate" UHECR acceleration sites can be studied in
detail in relation with the Auger signal, such as the giant radio lobes of
radio galaxies \cite{melia,berezhko}, the large scale jets
\cite{dermer}, or the AGN central engine. The only way
to study the acceleration mechanisms in the individual UHECR sources
is to single out the signal related to the UHECR production and
propagation in the broad band electromagnetic spectrum or to detect
the high-energy neutrino emission from the source.

Recent observations of fast-variability of the very-high-energy (VHE)
$\gamma$-ray emission from a nearby radio-galaxy M87
\cite{aharonian06} and from several blazars, Mkn 501 \cite{albert07},
PKS 2155-304 \cite{aharonian07} strengthen the conjecture that the
central engines of AGN, the supermassive black holes (BHs), could
operate as powerful particle accelerators
\cite{neronov07,krawczynski07,neronov08a}. The observed variability
time scales, which are of the order of, or shorter than the light
crossing times of the BH horizons, put tight constraints on the
possible locations of particle accelerators in these objects.

The compactness of the accelerators operating in the vicinity of the
supermassive BHs (the characteristic size scale is set up by the
Schwarzschild radius, $R_{\rm Schw}=2GM\simeq 3\times
10^{13}\left[M/10^8M_\odot\right]$~cm, where $M$ is the BH mass and $G$ is
the Newton's constant) makes particle acceleration in these objects
difficult, because of the inevitably strong energy losses related to
the synchrotron/curvature radiation of the accelerated particles (see e.g.
\cite{levinson00,aharonian02,neronov05}). The fact that, in spite of the strong
energy loss rate, these accelerators produce electrons with energies
of at least 10-100~TeV (the spectrum of the radio galaxy M87 extends
at least up to 20 TeV without a signature of a cut-off), implies that
the mechanism of acceleration operating in these objects is highly
efficient \cite{neronov07}. An obvious candidate for such a highly
efficient (i.e. fast) acceleration mechanism is acceleration of
charged particles in a strong large scale electric field, which can be
induced in the BH magnetosphere e.g. via the rotational drag of the
magnetic field by the black hole, or in the regions of magnetic field
reconnection in the accretion disk.

Large scale ordered electric fields, induced by rotation of the
compact object, are known to be responsible for particle acceleration
and high-energy radiation in pulsars (see e.g. \cite{lyne}). In the
case of pulsars, it is known that particles are accelerated in the
so-called "vacuum gaps" in magnetosphere, in which the
rotation-induced electric field is not neutralized by redistribution
of charges.  A similar mechanism of generation of large electric
fields can be realized in the vicinity of a rotating BH
\cite{wald,bicak}. Moreover, it has been argued that vacuum gaps
should form also in the vicinity of a rotating BH
\cite{lovelace76,lovelace79,blandford77,beskin92}, which means that a
mechanism of particle acceleration similar to the one operating in
pulsars should work also in the BH-powered sources.
The observational consequences of acceleration in the vacuum gaps of
BH magnetospheres were studied as a possible mechanism of powering the
AGN jet \cite{lovelace76,neronov02,neronov03} and neutrino emission
\cite{neronov02a} from AGN.

The possibility of UHECR generation near the supermassive
BH in the nuclei of normal galaxies (the so-called â"dead quasars"),
was studied in the Refs.  \cite{boldt,levinson00,neronov05}. As it is
mentioned above, in this particular case the compactness of the source
makes the acceleration to the UHECR energies ($>10^{20}$ eV) only
marginally possible.  A numerical study of the limit on the maximal
energies attainable by the accelerated protons reported in
Ref. \cite{neronov05} has revealed that, in general, acceleration to
the energies above $10^{20}$~eV is possible only in the vicinity of
the most massive BHs (with the masses of the order of
$10^{10}M_\odot$), which are relatively rear. Besides, such
acceleration is accompanied by strong electromagnetic emission, with a
luminosity about 2-3 orders of magnitude higher than the UHECR
luminosity of the source. This implies that the electromagnetic
luminosity of each of the UHECR sources should be at the level of
$10^{45-46}/N_{\rm source}$~erg/s, where $N_{\rm source}$ is the
number of the nearby UHECR sources (a typical distance to the source
$D\sim 50$~Mpc is assumed).  If the number of UHECR sources is not
very large, such a luminosity is about the luminosity of the quasars.
This, apparently, rules out the model of particle acceleration near
the supermassive BHs in the nearby normal galaxies (the "dead quasars"
\cite{boldt}) as a possible mechanism of UHECR production in the
nearby sources \cite{neronov05}.

The results of the Ref. \cite{neronov05} were obtained for the case of
acceleration of protons and under the assumption of a general
orientation of magnetic field (created by the matter accreting onto
the BH) with respect to the rotation axis of the BH. 

The assumption of the absence of alignment between the direction of
the magnetic field and the black hole rotation axis was justified in
the context of the "dead quasar" model of
environment of the supermassive BH.  On the contrary, steady AGN
activity in the source would result in the production of a stationary
accretion disk whose rotation axis is almost aligned with the rotation
axis of the BH \cite{bardeen-petterson}.  Such an alignment would lead
to the alignment of the magnetic field with the BH rotation axis.  In
this paper we investigate (both analytically and numerically) the
consequence of alignment of magnetic field with the BH rotation axis
for the particle acceleration.  We show that if the magnetic field is
aligned to within several degrees, production of ultra-high energy
protons becomes possible even for BH of moderate mass ($M\sim
10^8M_\odot$), provided that the magnetic field is very strong ($\sim
10^5$~G).  We also show that the constraint on the strength and
alignment of the magnetic field is largely relaxed in the case of
acceleration of heavy nuclei to the ultra-high energies ($\ge
10^{20}$~eV).

The assumptions of alignment of magnetic field with the BH rotation
axis and/or of acceleration of heavy nuclei lead also to a decrease of
the estimate of the power of electromagnetic radiation which
accompanies the acceleration.  In these cases the
electromagnetic luminosity of the vacuum gap at least does not exceed
the typical luminosities of the AGN in the local Universe.

The plan of the paper is as follows.  In Section
\ref{sec:magnetosphere} we discuss the possible location, geometry and
physical conditions in the vacuum gaps close to the BH horizon. To
make the presentation self-contained, we give the exact expressions
for the rotation induced electric field around a BH placed in an
external magnetic field inclined with respect to the rotation axis. In
Section \ref{sec:motion} we discuss the motion of a charged particle
in external gravitational and electromagnetic field near the black
hole. In Section \ref{sec:numeric} we present the results of numerical
modelling of such motion and find the dependence of the maximal
energies of protons accelerated by the rotation-induced electric field
on the main geometrical and physical parameters of the vacuum gap. We
find that for reasonable assumptions about the BH mass and the
strength of magnetic field the acceleration of protons to the UHECR
energies is possible only in the case when the magnetic field is
almost aligned with the rotation axis of the black hole (it remains to
be seen if such alignment can be achieved in Nature). In Section
\ref{sec:electromagnetic} we calculate, analytically and numerically,
the electromagnetic luminosity of the gap and find that the alignment
of the magnetic field with the rotation axis reduces the the \gr\
luminosity of the gap (making the estimate of the electromagnetic
power compatible with the estimates of the luminosity of typical AGN
in the local Universe). We find that, apart from the boosting of the
maximal energies of accelerated particles and reduction of the
electromagnetic power, the alignment of the magnetic field with the
rotation axis results in reduction of the probability of the onset of
pair production inside the gap which would lead to a ``discharge'' of
the gap and neutralization of the large scale electric field. In
Section \ref{sec:nuclei} we explore the acceleration of heavy nuclei
in the vacuum gaps and show that, contrary to the case of proton
acceleration, the energies of nuclei can reach $10^{20}$~eV already at
moderate magnetic field strength and in the absence of alignment of
the magnetic field with the rotation axis. Finally, in Section
\ref{sec:conclusions} we summarize our results.

\section{Vacuum gaps in the BH  magnetosphere}
\label{sec:magnetosphere}

\subsection{Black hole magnetosphere}  
 
 In the models of the most powerful AGN, accretion is assumed to
 proceed at nearly the maximal possible (Eddington) rate at which the
 gravitational attraction by the BH starts to be balanced by the
 pressure of radiation produced by the accretion flow. In this regime
 the accretion flow forms a geometrically thin, optically thick
 accretion disk \cite{shakura}, emitting thermal radiation, which, in
 the case of AGN, is usually identified with the so-called "big blue
 bump" in the ultra-violet part of the AGN spectrum. When the
 accretion rate onto the BH is sufficiently below the Eddington rate,
 the accretion proceeds in a "radiatively inefficient" regime, in
 which most of the released gravitational energy of the accreting
 matter is converted into the internal energy, rather than into the
 radiation. This energy is subsequently advected by the BH or ejected
 with a matter outflow \cite{narayan}. In this regime, large internal
 energy of matter does not allow formation of a gemetrically thin
 accretion disk. Instead, the accreting matter forms a hot, optically
 thin gaseous torus around the BH \cite{rees}.

In both accretion regimes, the structure of the accretion flow close
to the black hole is significantly affected by the presence of strong
magnetic field. The details of the structure of magnetic field play,
in fact, determining role in the dynamics of matter in the regions in
which the magnetic field energy density $U_B=B^2/8\pi$ dominates over
the kinetic energy density of the accreting matter $U_k=\rho v^2/e$,
i.e. in the regions where the magnetization parameter
$\sigma=U_B/U_k>1$. Even though the density of matter in the
magnetized regions is very low, it is usually supposed to be
sufficient to neutralize the large scale electric field component
parallel to the magnetic field lines, which otherwise would be induced
by the rotation of the magnetic field and of the space-time around the
black hole (see below).  The characteristic charge density needed to
neutralize the parallel component of electric field in the
magnetosphere of a BH of the mass $M$ rotating with an angular velocity $\vec\Omega$
placed in an external magnetic field $\vec B$ is the so-called
"Goldreich-Julian" density \cite{goldreich}
\begin{equation} 
\label{GJ} 
n_q\sim\frac{\vec\Omega\cdot\vec B_{\rm ord}}{e} 
\sim\frac{aB_{\rm ord}}{e(GM)^2}  
\end{equation} 
where $G$ is the gravitational constant, $e$ is the charge of electron and $a$ is the BH angular momentum per unit mass (we use the Natural Units $c=\hbar=1$ throughout the paper).
If the density of the plasma in the large magnetization regions is
above $n_q$, the plasma and magnetic fields in the high-magnetization
regions form the "force-free" magnetosphere (in which the electric
field is everywhere orthogonal to the magnetic field).

Charged particles can be supplied to the force-free magnetosphere in
several ways. First, if the accretion flow generates a poloidal
magnetic field penetrating through the accretion disk, particles bound
to the magnetic field could be accelerated by a centrifugal force and
escape along the poloidal magnetic field lines to the
high-magnetization regions and further into a directed outflow
\cite{blandford-payne} (this mechanism is often suggested as the
mechanism of launching of collimated jet from the BH). Injection of
plasma via such mechanism is possible only along the magnetic field
lines at which the centrifugal force dominates over the gravitational
force. This condition is not satisfied in the regions close to the
black hole horizon and/or close to the black hole rotation axis. This
can lead to the deficiency of supply of charged particles and
formation of "vacuum gaps" in the force-free magnetosphere.

Once the "force-free" conditions in the vacuum gap in BH magnetosphere
are broken, the presence of a non-zero component of electric field
parallel to the magnetic field lines starts to provide a further
obstacle for the charge supply in the magnetosphere, leading to a
growth of the vacuum gap size. The process of the growth of the vacuum
gap can terminate only if the large enough electric field in the gap
leads to its "discharge", similar to the discharge in a capacitor.

The discharge of the gap can provide an additional mechanism of supply
of charged particles into the magnetosphere. This mechanism is similar
to the one operating in pulsar magnetospheres: particles accelerated
in the vacuum gaps can generate $e^+e^-$ pairs via pair production in
interactions with the magnetic or radiation fields. Such a mechanism
of charge supply is implied e.g. in the Blandford-Znajek scenario,
\cite{blandford77}. The efficiency of this mechanism of charge supply
strongly depends on the strength and inclination of the magnetic field
(see section \ref{sec:pair} below). The efficiency of the charge
supply via the pair production by \gr s on the soft infrared
background depends on the compactness of the infrared source ($\propto
L_{\rm IR}/R$) \cite{neronov07}.

Strictly speaking, the geometry and the structure of electromagnetic
field in the vacuum gap in the magnetosphere can be derived only from
numerical modeling of the process of formation of the gap. Such a
numerical modeling is a complicated task, because the existing
numerical codes for modelling of the black hole magnetospheres are
based on the explicit assumption of the force-free conditions, which
simplify the numerical calculations \cite{komissarov}. In fact, the
numerical codes have an upper limit on plasma magnetization
above
which they fail. Because of this difficulty, the existing numerical
codes always include an artificial injection of free charges in the
points where the vacuum gaps should normally form \cite{komissarov}.

At the same time, some of the qualitative features of the particle
acceleration in the vacuum gaps could be established via analytical
/numerical calculations done in the approximation opposite to the
approximation of the force-free magnetosphere, i.e. neglecting the
possible non-zero particle density in the vacuum gap. In what follows
we adopt such an approximation, leaving the self-consistent
calculations of the vacuum gap geometry and particle acceleration in
the vacuum gap to the future work.

\subsection{Rotation-induced electric field.}
\label{sec:fields}

A rotating BH placed in an external magnetic field
generates an electric field of quadrupole topology via a mechanism
similar to the mechanism of generation of a quadrupole electric field
by a rotating dipole magnetic field in the case of a neutron star. The
only difference is that in the case of the black hole the rotation of
external magnetic field arises due to a rotational drag of external
magnetic field, rather than due to the rotation of the star.

An exact solution of Maxwell equations in the background of Kerr
space-time of a rotating BH is known for the case of arbitrary
inclination angle of an asymptotically homogeneous magnetic field
\cite{wald,bicak}. In this solution, the rotational drag of magnetic
field by the BH leads to the generation of a large scale electric
field close to the BH horizon.  This solution can be a first
approximation for description of the structure of electromagnetic
field in the vacuum gap in the black hole magnetosphere.  It is clear,
however, that the presence of charge-separated plasma around the gap
will, in general, lead to modification of the electromagnetic field
structure.

A rotating BH is described by two parameters: its mass $M$ and the
angular momentum per unit mass $a<M$ (we use the system of units in
which the Newton's constant $G_N$ and the speed of light $c$ are equal
to 1).  The geometry of space-time in the vicinity of the horizon is
described by the Kerr metric \cite{bardeen72}
\begin{equation}
\label{kerr}
ds^2=-\alpha^2dt^2+g_{ik}\left(dx^i+\beta^idt\right)\left(dx^k+\beta^kdt\right)
\end{equation}
\begin{equation}
\alpha=\frac{\rho\sqrt{\Delta}}{\Sigma};\ \ g_{rr}=\frac{\rho^2}{\Delta};\ \ 
g_{\theta\theta}=\rho^2;\ \ g_{\phi\phi}=\frac{\Sigma^2\sin^2\theta}{\rho^2};
\end{equation}
\begin{equation}
\beta_\phi=
-\frac{2aMr}{\Sigma^2}; \ \ \Delta=r^2+a^2-2Mr;
\end{equation}
\begin{equation}
\Sigma^2=(r^2+a^2)^2-a^2\Delta\sin^2\theta;\ \ 
\rho^2=r^2+a^2\cos^2\theta
\end{equation}
The horizon is situated at $r_H=M+\sqrt{M^2-a^2}$. 

The solution of the Maxwell equations which corresponds to the
asymptotically homogeneous magnetic field inclined at angle an $\chi$
with respect to the BH rotation axis is given by the electromagnetic
tensor \cite{bicak}
\begin{eqnarray}
\label{fmunu}
  F_{tr}&=& \frac{aMB_0}{\rho^4\Delta}\left[\cos\chi
\Delta (r^2-a^2\cos^2\theta)(1+\cos^2\theta)+\right.\nonumber\\ &&\left.
\sin\chi r\sin\theta\cos\theta
\left\{
(r^3-2Mr^2+
ra^2(1+\sin^2\theta)+
2Ma^2\cos^2\theta)\cos\psi-\right.\right.\nonumber\\ &&\left.\left.
a(r^2-4Mr+a^2(1+\sin^2\theta))
\sin\psi\right\}\right]\nonumber\\       
     F_{t\theta}&=&  \frac{aMB_0}{\rho^4}\left[
2\cos\chi r\sin\theta\cos\theta(r^2-a^2)+\right.\nonumber\\ &&\left.\sin\chi(r^2\cos 2\theta+a^2\cos^2\theta)(a\sin\psi-r\cos\psi)
\right]
\nonumber\\
     F_{t\phi}&=& \frac{B_0\sin\chi aM}{\rho^2}\sin\theta\cos\theta(a
\cos\psi+r\sin\psi)\nonumber\\       
     F_{r\theta}&=& -B_0\sin\chi(a\cos\psi+r\sin\psi)-\nonumber\\
&&\frac{B_0\sin\chi a}{\Delta}
\left[(Mr-a^2\sin^2\theta)
\cos\psi-a(r\sin^2\theta+M
\cos^2\theta)\sin\psi\right]\nonumber\\       
     F_{r\phi}&=& B_0\cos\chi r\sin^2\theta+a\sin^2\theta F_{tr}-\nonumber\\ &&B_0\sin\chi\sin\theta
\cos\theta
\left[\left(r-a^2M/\Delta\right)
\cos\psi-a\left(1+rM/\Delta\right)\sin\psi\right]\nonumber\\     
     F_{\theta\phi}&=& B_0\cos\chi\Delta\sin\theta\cos\theta+\nonumber\\ &&
\frac{(r^2+a^2)}{a}F_{t\theta}+
B_0\sin\chi\left[(r^2\sin^2\theta+Mr\cos 2\theta)
\cos\psi-\right.\nonumber\\&&\left.
a(r\sin^2\theta+M\cos^2\theta)\sin\psi\right]
\end{eqnarray}
where 
\begin{eqnarray}
\psi&=&\phi+\frac{a}{2\sqrt{M^2-a^2}}\ln\left[\frac{r-M+\sqrt{M^2-a^2}}
{r-M-\sqrt{M^2-a^2}}\right].
\end{eqnarray}

In order to understand the qualitative features of an electromagnetic
field configuration in a curved space-time (like the one of the
rotating BH) it is convenient to consider the field geometry in a
locally Lorentzian reference frame. In the case of Kerr space-time it
is convenient to choose the so-called ``locally non-rotating''
reference frame (LNRF), spanned by the orthonormal basis vectors
\cite{bardeen72}
\begin{eqnarray}
\label{zamo}
e_{\hat 0}&=&\frac{\Sigma}{\rho\sqrt{\Delta}}\frac{\partial}{\partial t}+
\frac{2Mar}{\Sigma\rho\sqrt{\Delta}}\frac{\partial}{\partial \phi};\nonumber\\
e_{\hat r}&=&\frac{\sqrt{\Delta}}{\rho}\frac{\partial}{\partial r};\nonumber\\   
e_{\hat\theta}&=&\frac{1}{\rho}\frac{\partial}{\partial\theta};\nonumber\\   
e_{\hat\phi}&=&\frac{\rho}{\Sigma\sin\theta}\frac{\partial}{\partial\phi}.
\end{eqnarray}
In this reference frame the magnetic and electric field vectors are
expressed through the components of the tensor of electromagnetic
field $F_{\mu\nu}$, given by Eq. (\ref{fmunu}), as
\begin{eqnarray}
     B^{\hat r}   &=& \frac{F_{\theta\phi}}{\Sigma\sin\theta};\ \ 
     B^{\hat \theta}  =-\frac{\sqrt{\Delta}F_{r\phi}}{\Sigma\sin\theta};\ \ 
     B^{\hat \phi}  = \frac{\sqrt{\Delta}F_{r\theta}}{\rho^2}
\end{eqnarray}
\begin{eqnarray}
     E^{\hat r}   &=& \frac{\Sigma F_{tr}-arF_{r\phi}/\Sigma}{\rho^2};\ \ 
     E^{\hat \theta}  = \frac{\Sigma F_{t\theta}-ar F_{\theta\phi}/\Sigma}{
       \rho^2\sqrt{\Delta}};\ \ 
     E^{\hat \phi}  = \frac{F_{t\phi}}{\sqrt{\Delta}\sin\theta}
\end{eqnarray}

\begin{figure}
\begin{center}
\includegraphics[width=0.7\linewidth]{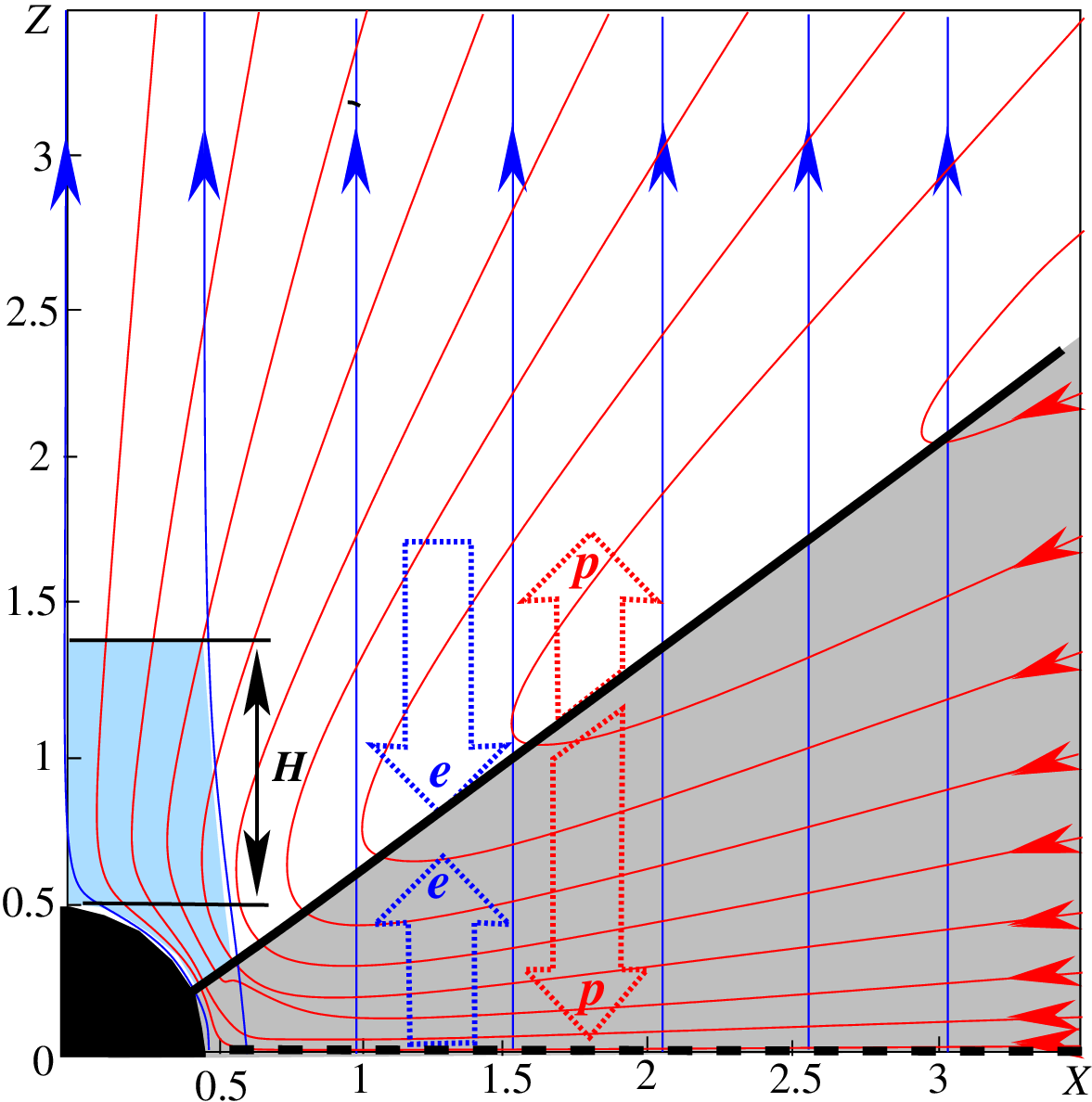}
\caption{Magnetic (blue solid) and electric (red solid) field lines
  of the field (\ref{fmunu}) for $a=M$ and $\chi=0$. The horizon is
  situated at $r=0.5(R_{grav})=M$. Wide blue and red arrows show the
  directions of acceleration of electrons and protons in different
  regions. Thick solid and dashed lines show the conical and
  equatorial ``force-free'' surfaces. Blue-shaded region shows the
  location of a vacuum gap into which particles from the equatorial
  accretion flow can leak if $\chi>0$.}
\label{fig:fields}
\end{center}
\end{figure}

The expressions for electric and magnetic field simplify in the case
of a magnetic field aligned with the BH rotation axis, $\chi=0$. The
solution of Maxwell equations for this particular case was initially
reported in the Ref. \cite{wald}. Fig.  \ref{fig:fields} shows the
geometry of electric and magnetic field lines for $\chi=0$ in the case
of a maximally rotating BH $a=M$.  Taking the asymptotic expression
for $\vec E, \vec B$ in the limit $r\rightarrow\infty$ one can find
that far away from the horizon the electromagnetic field is
\begin{eqnarray} 
B^{\hat r}&\rightarrow& B_0\cos\theta;\ \ 
B^{\hat\theta}\rightarrow -B_0\sin\theta\nonumber\\
E^{\hat r}&\rightarrow& -\frac{B_0aM(3\cos^2\theta-1)}{r^2};\ \ 
E^{\hat\theta}=O(r^{-4}).  
\label{eq:asymptotic}
\end{eqnarray}
The electric field has a quadrupole geometry, which is clearly visible
in Fig \ref{fig:fields}.
 
\begin{figure}
\begin{center}
\includegraphics[width=\linewidth]{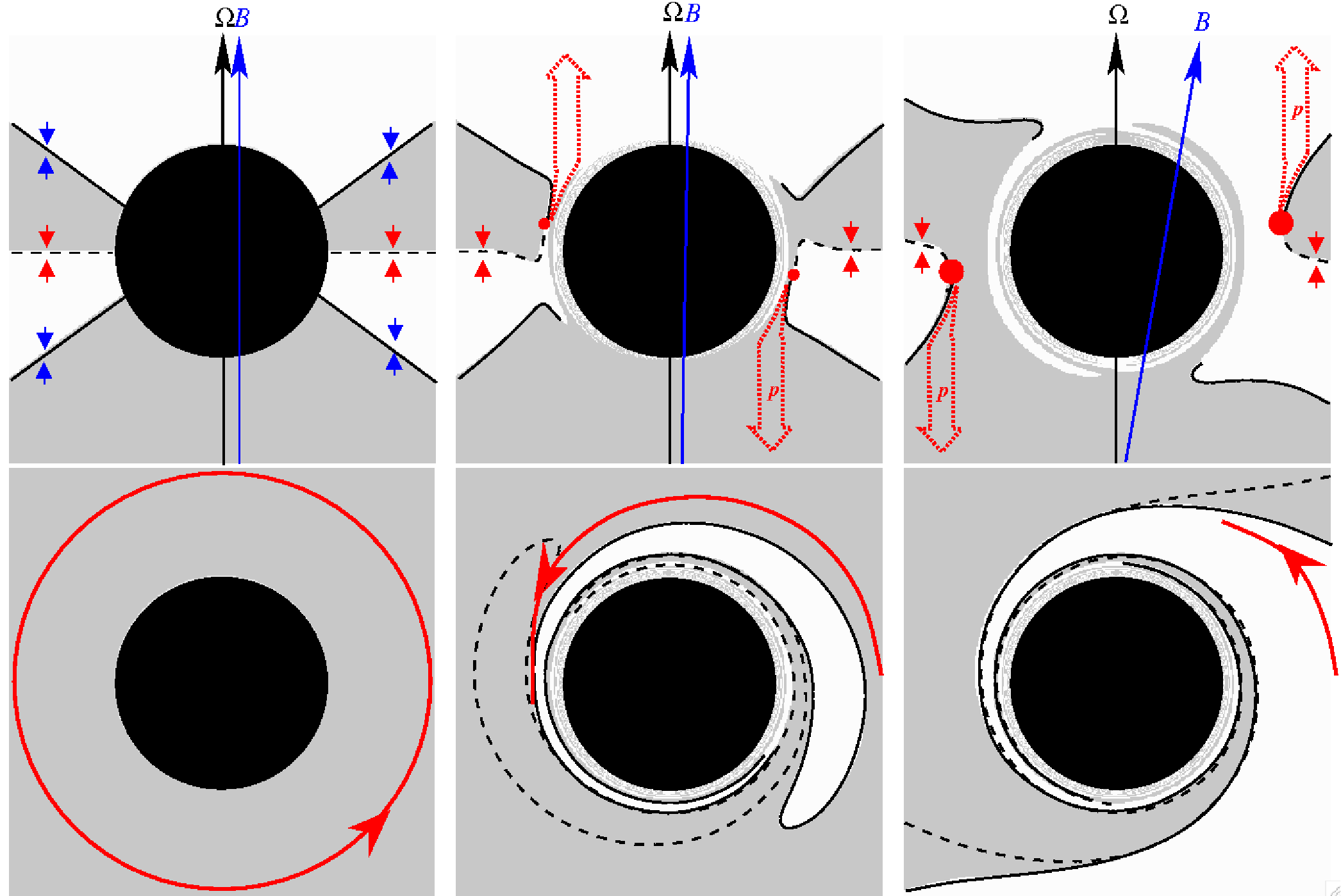}
\caption{Evolution of the structure of the force-free surfaces with
  the increasing misalignment of the magnetic field. The equatorial
  force-free surface, shown by the dashed line, "reconnects" with the
  conical force-free surfaces (solid lines) at non-zero inclination
  angles. Three top panels show the view in the $xz$ plane for the
  inclination angles $\chi=0^\circ,\chi=1^\circ$ and
  $\chi=10^\circ$. The bottom panels show the view in the $xy$ plane
  (just above the equatorial plane, at $z=0.01R_{\rm Schw}$) for the
  same inclination angles.  The spiral-like reconnection surfaces,
  which would be visible if the plane $z=-0.01R_{\rm Schw}$ would be
  chosen, are shown by the black dashed curves in the middle and right
  lower panels.  White (grey) areas show the regions where the
  electric field component along the magnetic field lines has positive
  (negative) sign. Black circles show the horizon of a maximally
  rotating BH. Depending on the charge sign, particles in-spiralling
  toward the BH along the equatorial force-free surface are either
  ejected to infinity (positively charged particles in this figure,
  whose typical trajectories are shown by red curves) or are trapped
  in the region between two force-free surfaces.}
\label{fig:forcefree}
\end{center}
\end{figure}

In the regions where electric field has a component along the magnetic field,
charged particles can be accelerated. Particles of opposite charge sign are
accelerated in opposite directions. The acceleration can be avoided only if a
particle resides at a surface at which the electric field is orthogonal to the
magnetic field (which can be called the ``force-free'' surfaces), such as the
surfaces shown by thick solid and dashed black lines in Fig. \ref{fig:fields}.
The force-free surfaces separate different acceleration regions (with electric
field directed along or oppositely to the magnetic field), shown as  white
($\vec E\cdot \vec B>0$) and grey ($\vec E\cdot \vec B<0$) areas in Figs.
\ref{fig:fields}, \ref{fig:forcefree}.

As one can see from Fig. \ref{fig:fields}, the force-free surfaces
serve as ``attractors'' to the particles of a definite charge sign. In
the case of magnetic field aligned with the rotation axis, shown in
Fig. \ref{fig:fields}, the surface attracting the positively charged
particles (the equatorial plane surface) does not intersect with the
force-free surface attracting negatively charged particles (shown by
the thick solid black curve). However, as soon as the inclination
angle of the magnetic field becomes non-zero, the two force-free
surfaces ``reconnect'' as it is shown in Fig. \ref{fig:forcefree}.  At
moderate inclination angles of the magnetic field, the reconnection
region is situated close to the equatorial plane, near the BH. The
width of the reconnection region grows with the increasing inclination
angle. In the following sections we show that charged particles which
spiral toward the BH in the equatorial plane, can be accelerated and
ejected to infinity when they reach the reconnection region.

\subsection{Geometry and location of the vacuum gap(s).}
\label{subsec:geometry}

In the electromagnetic field configuration around the BH discussed in the
previous section the vacuum gap forms in the region above the BH
``polar cap'', if the free charges which can fill the magnetosphere
are supplied by the equatorial accretion disk. Extraction of the
free charges from the equatorial disk by the rotation-induced electric
field will result in the formation of a region filled with electrons,
above and below the equatorial disk, up to the conical force-free
surface, shown by the black solid thick curve in this figure. However,
if the magnetic field is aligned with the rotation axis, the region
above the ``polar cap'' of the BH will remain free of charge, because
there is no way to fill this region via extraction of charges from the
equatorial disk.  Only if the magnetic field is mis-aligned with the
BH rotation axis, the free charges from the equatorial disk can
``leak'' into the polar cap passing through the ``reconnection''
region of the two force-free surfaces, as it is shown in Fig.
 \ref{fig:forcefree}. This will result in charge supply into the
region above the ``polar cap'', which is shown as a blue-shaded region 
in
Fig. \ref{fig:fields}. However, any charged particle which penetrates
into this region, escapes to infinity along the magnetic field lines,
so that the charge distribution in this region should be constantly
``refilled'' via the ``leakage'' from the equatorial force-free
surface.

In principle, the
re-distribution of the charges supplied from the accretion disk will
modify the structure of the electromagnetic field around the BH. The
exact location of the vacuum gap(s) in the BH magnetosphere can be
found only via numerical modeling of the charge supply and formation
of the force-free magnetosphere. Such modeling is difficult already in
the case of pulsars, where several ``candidate'' sites for the vacuum
gap locations are considered, such as the magnetic polar cap regions,
``slot gaps'' at the boundaries of regions filled with positively and
negatively charged plasma, ``outer gaps'' close to the light cylinder
etc (see e.g. \cite{lyne}). In the case of the black holes, modeling
of the formation of the force-free magnetosphere has an additional
difficulty, because the location of the gap can depend on the
geometrical pattern of the accretion flow.  The gap model
considered in the previous section can serve as a useful ``toy
model'' for the gap geometry.

Within this "toy model" the gap has a particular geometry shown in Fig.
\ref{fig:fields}.  Charged particles "leak" from the  accretion disk through
the region of "reconnection" of the force-free surfaces (see Fig.
\ref{fig:forcefree}). The gap is situated above the BH ``polar cap'' and extend
along the magnetic field lines up to the height $H$ above the BH horizon.  We
consider particle acceleration in a gap of such geometry, to establish
qualitative features of the vacuum gap acceleration mechanism. Our results can
be then generalized, in a straightforward way, to the more complicated (and
more realistic) models of vacuum gaps in the black hole magnetosphere.

\section{Particle acceleration in the vacuum gap}
\label{sec:motion}

\subsection{General properties of particle motion in the vacuum gap}
\label{subsec:properties}

The electromagnetic field in the vacuum gap consists of electric and
magnetic field which are, in general, inclined at a non-zero angle
with respect to each other. General properties of particle motion in
such electromagnetic field and in the curved space-time around the
BH can be readily understood.

Qualitatively, a particle moving in a magnetic
field spirals around the magnetic field lines with a gyro-radius
\begin{equation}
\label{giro}
R_{gyro}=\frac{\it E}{qB_0}\approx 10^{2}\left[
\frac{10^4\mbox{ G}}{B_0}\right]
\left[\frac{{\it E}}{1\mbox{ GeV}}\right]\mbox{ cm}
\end{equation}
where ${\it E}$ is the energy of the particle and $q$ is its charge.
For a particle with an energy ${\it E}\ll 10^{20}$~eV the gyro-radius
is normally essentially less than the scale of variation of the
electromagnetic field which is of order of $R_{\rm Schw}=3\times
10^{13}\left[M/10^8M_\odot\right]$~cm. This gyro-radius distance scale
is also much smaller than the scale at which particle trajectory is
significantly affected by the gravitational field of the black hole
(which is also about $R_{\rm Schw}$). Therefore, one could separate
the distance scales at which electromagnetic and gravitational forces
are important. At small scales ($\ll R_{\rm Schw}$) one can locally
approximate particle trajectories as motion in the "crossed" magnetic
and electric fields, forming the field configuration described in the
Section \ref{sec:fields}.

The motion of a charged particle in such electromagnetic field
is composed of the spiraling along the magnetic field lines and the 
orthogonal drift with velocity 
\begin{equation}
v_{drift}=\frac{\vec E\times \vec B}{|\vec B|^2}
\end{equation}
Since in the case of interest $|E|\sim |B|$, $v_{drift}\sim c$ and, in
general, particle trajectories significantly deviate from a simple
spiraling along the magnetic field lines. For example, the drift
velocity of particles moving in the equatorial plane near the black
hole placed in the magnetic field, aligned with the black hole
rotation axis, is directed along a circle around the black hole, so
that the particles drift along the circular trajectories (red circle
in the left bottom panel of the Fig. \ref{fig:forcefree}). If the
magnetic field is slightly mis-aligned with the rotation axis, the
particles spiral into (or away from) the black hole, rather than move
along the circular orbits (red curves in the middle and right bottom
panels of the Fig. \ref{fig:forcefree}). The particles, which spiral
in, reach the region of ``reconnection'' of the force-free surfaces
(red dots in Fig. \ref{fig:forcefree}) and are finally either ejected
into the ``polar cap'' region or fall under the BH horizon.

Apart from spiraling along the magnetic field lines and drift across
the field lines, particles can be accelerated by the electric
field. Only the presence of electric field component along the
magnetic field line can lead to the increase or decrease of particle
energy. Obviously, positively charged particles are accelerated in the
direction opposite to the direction of acceleration of negatively
charged particles. Depending on the location of the point of injection
of a low-energy particle (e.g. proton or electron) into the
reconnection region, the particle can be accelerated in such a way
that it either escapes ``to infinity'' along the direction of magnetic
field lines, or it is trapped in a finite region between the
neighbouring force-free surfaces close to the horizon.

\subsection{Maximal energies of accelerated particles}
\label{sec:maximal}

Neglecting the energy losses, the maximal energies attainable for
nuclei of the charge $Ze$ are determined by the available potential
difference in the gap, $U\sim ER\sim aBR$,
\begin{equation}
\label{potential} 
{\it E}_{\rm max}= ZeBR\simeq 10^{20}Z \left[{B\over 10^4\, \mbox{G}}\right]
\left[\frac{M}{10^{8}M_\odot}\right]\mbox{ eV.}
\end{equation} 
In a realistic situation the energy losses of the accelerated
particles (e.g. on the electromagnetic radiation associated with the
accelerated motion of the particle) limit the maximal energies to the
values below the estimate of Eq. (\ref{potential}). 

In particular, the accelerated nuclei inevitably suffer from the  curvature
radiation loss
\begin{equation}
\frac{d{\it E}}{dt}=-\frac{2Z^2e^2{\it E}^4}{3A^4m_p^4R^2},
\label{eq:cur}
\end{equation}
where $m_p$ is the proton mass and $A$ is the atomic number. In the
electromagnetic field configuration (\ref{fmunu}), describing a
magnetic field almost aligned with the BH rotation axis, the curvature
radius of magnetic field lines in the gap scales roughly as $R\simeq
R_{\rm Schw}/\chi$.  Equating the acceeration rate $dE/dt\sim ZeB$ to minus the energy loss rate (\ref{eq:cur}), one appears at the following estimate of the maximum
energy,
\begin{eqnarray}
{\it E}_{\rm cur}&=&\left[{3A^4m^4 R^2 B \over 2Ze\chi^2}\right]^{1/4}\simeq \nonumber\\
&&8\times
10^{19}A Z^{-1/4}\left[\frac{M}{10^{8}M_\odot}\right]^{1/2}
\left[\frac{B}{10^4\mbox{ G}}\right]^{1/4}\left[\frac{\chi}{1^\circ}\right]^{-1/2}\mbox{ eV}.
\label{curv_cut}
\end{eqnarray}
The range of applicability of Eq. (\ref{curv_cut}) is
given by the  condition $B>B_{\rm crit}$, where $B_{\rm crit}$ is found from the requirement ${\it E}_{\rm cur}={\it E}_{\rm max}$:
\begin{equation}
B_{\rm crit} = \left[{3A^4m_p^4 \over 2Z^5e^5R^2\chi^2}\right]^{1/3} 
\simeq 10^4 A^{4/3}Z^{-5/3}\left[{M\over 10^{8}M_{\odot}}\right]^{-2/3}
\left[\frac{\chi}{1^\circ}\right]^{-2/3} \;{\rm G}.
\label{eq:Bcrit}
\end{equation}
This critical field corresponds to particle energy 
\begin{equation}
\label{ecrit}
{\it E}_{\rm crit}\approx 10^{20}A^{4/3}Z^{-2/3}
\left[{M\over 10^{8}M_{\odot}}
\right]^{1/3}\left[\frac{\chi}{1^\circ}\right]^{-2/3}
\mbox{ eV}, 
\end{equation} 
 
The above estimates show that it is possible to accelerate protons or
heavier nuclei to the energies about $10^{20}$~eV in the central
engines of typical AGNs, with the BH masses around $10^8$
solar masses. The necessary condition for such acceleration is that
the curvature radius of particle trajectories should be much larger
(by a factor $1/\chi$) than $R_{\rm Schw}$. This can be achieved in the
``polar cap'' regions, if the magnetic field is moderately misaligned with
the BH rotation axis.

\section{Numerical modeling of particle acceleration in the gap}
\label{sec:numeric}

In order to study the dependence of the maximal energies of the
accelerated particles on the parameters of the model (in particular,
on the BH mass $M$, magnetic field $B$ and 
inclination angle $\chi$ of the magnetic field with respect
to the BH rotation axis), we have developed a numerical code
which traces the particle trajectories through the electromagnetic
field configuration described in Section \ref{sec:magnetosphere}
 with the account of the
energy loss on the emission of synchrotron/curvature radiation. To
study the electromagnetic luminosity of the vacuum gap we also model
the propagation of photons emitted by the accelerated particles
through the BH space-time.

\subsection{Numerical code}

We have developed a numerical code to model charged particle motion in
the curved space-time and in the electromagnetic field configuration
described Section \ref{sec:fields}. 
For this we have first written the
equations of motion of a charged particle in the LNRF (see
Eq. \ref{zamo}) \cite{thorne86}
\begin{equation} 
\label{ham2} 
\frac{d{\vec p}}{d\hat t}= 
e(\vec E+\vec v\times \vec B)+m\gamma\vec g+\hat H\vec p+\vec f_{rad} 
\end{equation} 
where $\vec p$ is the particle momentum 
\begin{equation} 
\label{momentum} 
p^{\hat a}=m\gamma v^{\hat a} 
\end{equation}
(hats denote the vector components in the LNRF) ${\vec g}$ is the
gravitational acceleration and ${\hat H}$ is the tensor of
gravi-magnetic force. The force ${\vec f}_{rad}$ is the radiation
reaction force and $v^{\hat a}$ is the 3-velocity of the particle in
the LNRF.  Next, taking into account the fact that the motion of a
charged particle involves many distance scales, from the gyroradius
$R_{gyro}$ (\ref{giro}) up to $R_{\rm Schw}$, we have found that a
step of the numerical integration is determined by the smallest scale
involved ($R_{gyro}$ most of the time). At this length scale the
motion of the particle can be modelled in the LNRF by a numerical
integration of Eqs. (\ref{momentum}). Using the book
\cite{Numerical_Recipes}, we have written a code, which integrates the
above equations numerically. The elementary step of the numerical
integration increases or decreases along the particle trajectory,
together with the increase or decrease of the energy of the particle
(see Eq. \ref{giro}).
 
The radiation reaction force $\vec f_{rad}$  
for the ultra-relativistic particles moving in external electromagnetic field 
is (see, e.g. \cite{landau}) 
\begin{equation} 
\label{reaction} 
\vec f_{rad} 
=\frac{2e^4\gamma^2}{3m^2}\left((\vec E+\vec v\times \vec B)^2-(\vec v\cdot(\vec E+\vec v\times \vec B))^2\right)\frac{\vec v}{|v|} 
\end{equation}  
Note that if particles move at large angle with respect to the
magnetic field lines, this expression will describe mostly synchrotron
energy loss. However, in the case when particles move almost along the
magnetic field lines, the last equation will "mimic" the effect of
curvature energy loss.

We follow the particle trajectories during the phase of in-spiralling
along the equatorial plane and subsequent propagation inside the
vacuum gap. The particles are supposed to be initially at rest, with
respect to the LNRF. Particles are initially injected in the
equatorial plane (to mimic the injection from the thin accretion
disk).

To study the electromagnetic emission from the accelerated particles
we calculate, at each step of the integration of particle trajectory,
the energy and power of synchrotron/curvature emission and ascribe
this energy and power to a nominal "photon" which is emitted along the
direction of the particle motion. We subsequently trace the trajectory
of each emitted photon from the emission point to infinity by
integrating the equations of the geodesics of the Kerr metric.

Charged particles which reach the outer boundary of the gap, escape
along the geodesics of the Kerr metric. The motion along the time-like
geodesics o the Kerr metric is determined by the equations
\cite{carter}
\begin{eqnarray}
\label{eq_motion}
r^2\dot r&=&\sqrt{T^2-\Delta(\mu^2r^2+(\Phi-aE)^2)}\nonumber\\
r^2\dot \phi&=&\Phi-aE+aT/\Delta\nonumber\\
r^2\dot t&=&a\Phi-a^2E+(r^2+a^2)T/\Delta
\end{eqnarray}
where $E$ and $\Phi$ are the integrals of motion, which correspond to
the conserved energy and angular momentum, $\mu$ is particle mass and
$T$ is defined as $T=E(r^2+a^2)-\Phi a$. In the following, the cited
energies of particles ejected from the gap are always the "energies
measured by an observer at infinity", i.e. the integrals of motion
$E$, which enter the geodesic equations (\ref{eq_motion}).

\subsection{Dependence of the maximal energy  on the gap parameters}
\label{sec:param}

In this section we study dependence of maximum energy to which protons
can be accelerated on the geometry of the gap, namely, on the basic
parameters of the model: the gap height $H$, the inclination of
magnetic field with respect to the BH rotation axis, $\chi$, the BH
mass $M$ and on magnetic field strength $B$.

\begin{figure}
\begin{center}
\includegraphics[width=0.75\linewidth]{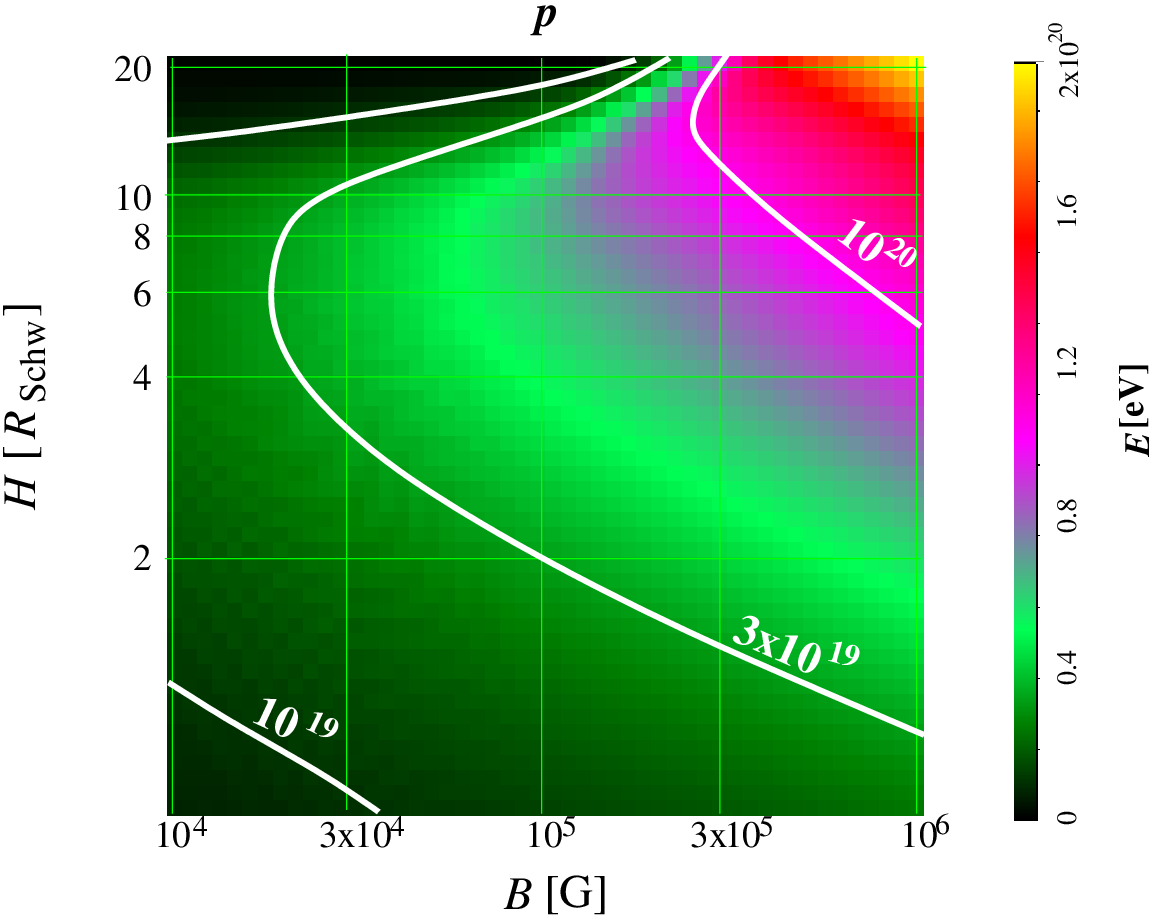}
\caption{Energies of protons ejected from the vacuum gap (measured by an
observer at infinity), as a function of the gap height $H$ and magnetic field 
$B$. The parameters 
are: BH mass $M=3\times 10^8M_\odot$, rotation moment per unit
mass $a=0.99GM$, the inclination angle of magnetic
field $\chi=5^\circ$. Protons are initially injected in the equatorial plane
at the distance $R=1.4r_H$ from the BH.}
\label{fig:E_vs_H}
\end{center}
\end{figure}

Fig.~\ref{fig:E_vs_H} shows the dependence of maximum energy of
accelerated protons on the gap height $H$. We set the inclination angle of
magnetic field $\chi=3^\circ$. One can see that for the assumed BH
mass, $M_{\rm BH}=3\times 10^8M_\odot$, the acceleration to the
energies above $10^{20}$~eV is possible only if the gap height is
larger than $H> R_{\rm Schw}$, provided that the magnetic field near
the horizon is extremely strong, $B>3\times 10^4$~G.  One can notice
that the increase of the gap height beyond a certain field-dependent
limit does not lead to the further increase of the particle
energies. This is explained by the decrease of the electric field
strength (see Eq.  \ref{eq:asymptotic}) and, as a consequence, the
decrease of the acceleration rate far from the BH. Since the magnetic
field is assumed to be asymptotically constant, the rate of the energy
loss does not decrease with the distance, contrary to the acceleration
rate. The combination of the decreasing acceleration rate and steady
loss rate leads to the decrease of the particle energies at large
distances. Obviously, the assumption of the asymptotically constant
magnetic field, adopted in our toy model, does not hold in a realistic
case, when the magnetic field is produced by the accreting matter.

\begin{figure}
\begin{center}
\includegraphics[width=0.75\linewidth]{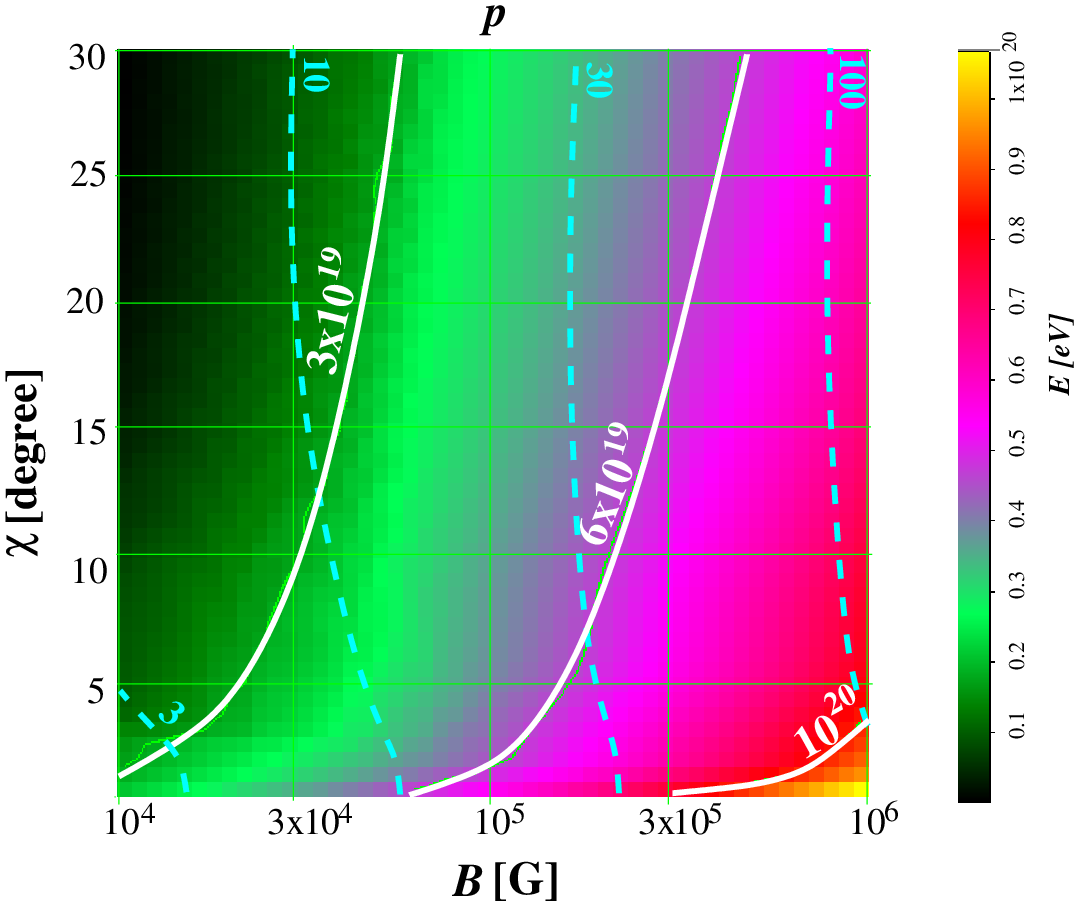}
\caption{ Energies of protons ejected from the vacuum gap (measured
  by an observer at infinity), as a function of the inclination angle
  of magnetic field, $\chi$ and of the magnetic field strength
  $B$. The parameters of the calculation are: $M=3\times 10^8M_\odot$,
  $a=0.99GM$, $H=5R_{\rm Schw}$. Dashed
  cyan contours show the ratio of the electromagnetic to cosmic ray
  luminosity for the particular gap parameters.}
\end{center}
\label{fig:E_vs_chi}
\end{figure}

Fig.~\ref{fig:E_vs_chi} shows the dependence of maximum energy on the
inclination angle of magnetic field with respect to the rotation axis
of the black hole, $\chi$. In our toy model, particles
are initially supplied by the equatorial accretion disk, so that the
initial positions of the particles are in the equatorial plane. 
One can see that the
acceleration to the energies above $10^{20}$~eV close to a BH of the
mass $5\times 10^8M_\odot$ is possible only if $\chi\ll 1$. The
increase of the misalignment between the magnetic field and the BH
rotation axis leads to the decrease of the maximal energies and to the
increase of the power of electromagnetic emission which accompanies
the acceleration process ((cyan dashed curves in this Figure are marked by
the value of the ratio of the electromagnetic luminosity, $L_{\rm EM}$
to the UHECR luminosity, $L_{\rm UHECR}$).

 The
requirement on the alignment of the magnetic field can be somewhat
relaxed in the case of higher black hole mass or still higher
magnetic field strength. In fact, since, in the "energy loss saturated"
regime the maximal energy grows as $B^{1/4}M^{1/2}$ (see
Eq. \ref{curv_cut}), the increase of the magnetic field by an order of
magnitude results just in a slight correction of the maximal allowed
misalignment angle, while an order of magnitude heavier black hole
accelerates particles to just a factor of 3 higher energies.

\begin{figure}
\begin{center}
\includegraphics[width=0.75\linewidth]{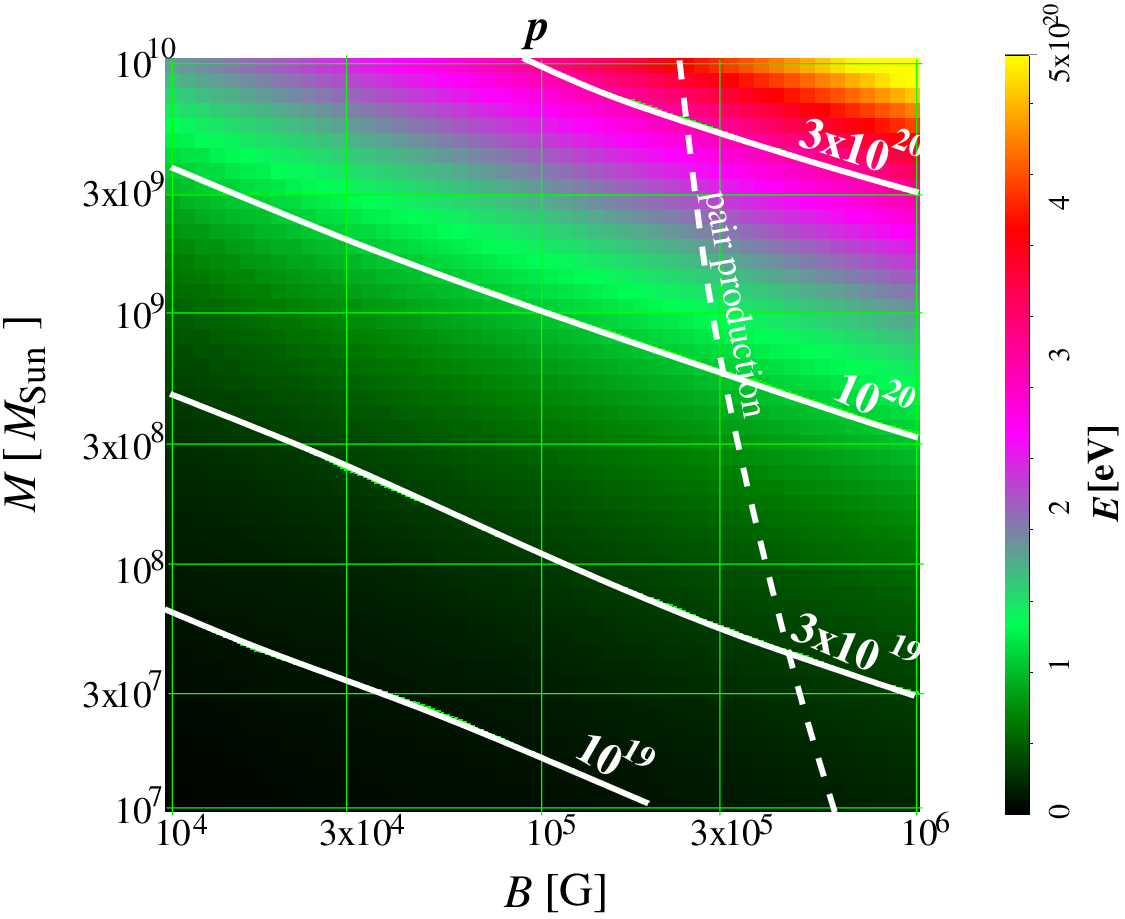}
\caption{Maximal energy of protons ejected from the vacuum gap (measured by an
observer at infinity), as a function of the BH mass $M$ and magnetic field $B$. 
The parameters  are:  $a=0.99GM$, 
$\chi=5^\circ$, $H=5R_{\rm Schw}$.
Particles are initially injected in the equatorial plane
at the distance $R=1.4r_H$ from the BH.}
\label{fig:E_vs_MB}
\end{center}
\end{figure}

This is clear from Fig.~\ref{fig:E_vs_MB} in which the dependence of
the particle energies on the magnetic field and the BH mass is
shown. From this Figure one can see that at the energies $E\ge
10^{20}$ eV the relation (\ref{curv_cut}) holds, so that at the levels
$E=const$ one has $M\sim B^{-1/2}$. For a $10^9M_\odot$ BH the energy
$10^{20}$ eV is attainable already when the magnetic field is $B\sim
10^4$~G. However, the further increase of the magnetic field by two
orders of magnitude, up to $B\sim 10^6$~G, results in an increase of
the particle energies just by a factor of $2$.

\section{Electromagnetic emission from the gap}
\label{sec:electromagnetic}

\subsection{Direct \gr\ emission from the particle acceleration in the gap}
 
Protons accelerated near the BH horizon in the gap can produce \gr\
emission in the VHE band through several radiation mechanisms.  For
example, TeV emission can be synchrotron or curvature \gr\ emission
which accompanies proton acceleration \cite{levinson00,neronov05}. The
energy of curvature photons produced by protons accelerated to the
energy $E_{\rm curv}$, given by Eq. (\ref{curv_cut}), is and
\begin{equation} 
\label{curvp} 
\epsilon_{{\rm curv},p}\simeq 2\left[\frac{E_{\rm curv}}{10^{20}\mbox{ eV}} 
\right]^3\left[\frac{M}{3\times 10^{8}M_\odot}\right]^{-1}
\left[\frac{\chi}{1^\circ}\right]\mbox{~TeV} 
\end{equation} 

\begin{figure}
\begin{center}
\includegraphics[width=0.75\linewidth]{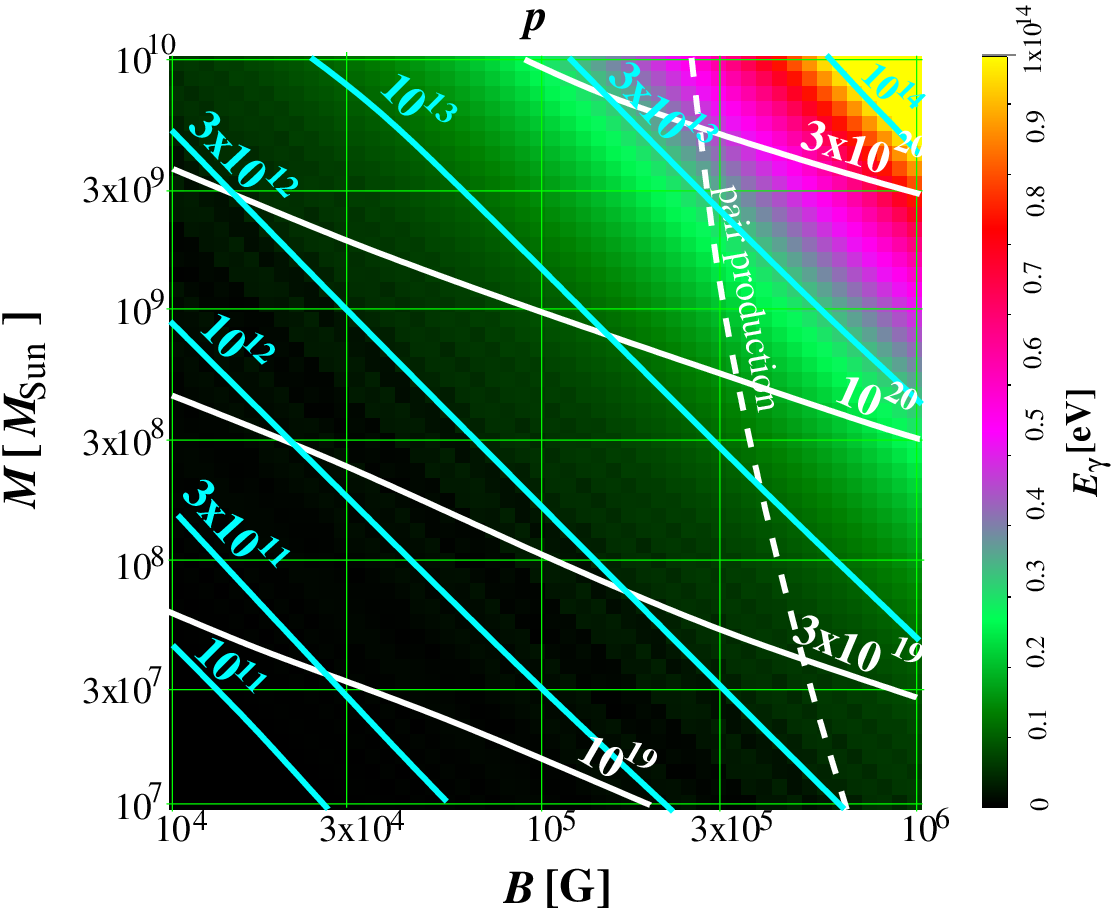}
\caption{Color and cyan contours: maximal energy of gamma quanta
  produced by the accelerated protons in the vacuum gap (measured in
  the LNRF), as a function of the BH mass and magnetic field. White
  contours show the maximal energies of the accelerated protons (see
  Fig. \ref{fig:E_vs_MB}).  Dashed line shows the values of $M,B$ at
  which the pair production in the magnetic field sets on on the
  distance scale of the inhomogeneity of the magnetic field. The
  parameters are: $a=0.99GM$, $\chi=5^\circ$, $H=5R_{\rm Schw}$.
  Protons are initially injected in the equatorial plane at the
  distance $R=1.4r_H$ from the BH.}
\label{fig:E_vs_g}
\end{center}
\end{figure}

Thus, particle acceleration activity of the gap should reveal
itself through the \gr\ emission in the very-high-energy (VHE)
band. Fig. \ref{fig:E_vs_g} shows numerically calculated dependence of the maximal energies of the
\gr s emitted by the gap on the physical parameters of the gap, $M,B$. One can
see that for the range of parameters at which UHECR production is possible
(i.e. above the thick white line with the  mark "$10^{20}$" in this figure) 
the maximal energies of the \gr s reach $1-100$~TeV. Combining Eqs.
\ref{curv_cut} and \ref{curvp} one can find that $\epsilon_{{\rm curv},p}\sim
M^{1/2}B^{3/4}$ so that the lines $\epsilon_{{\rm curv},p}=const$ correspond to
$M\sim B^{-3/2}$ in the figure.

\begin{figure}
\begin{center}
\includegraphics[width=0.75\linewidth]{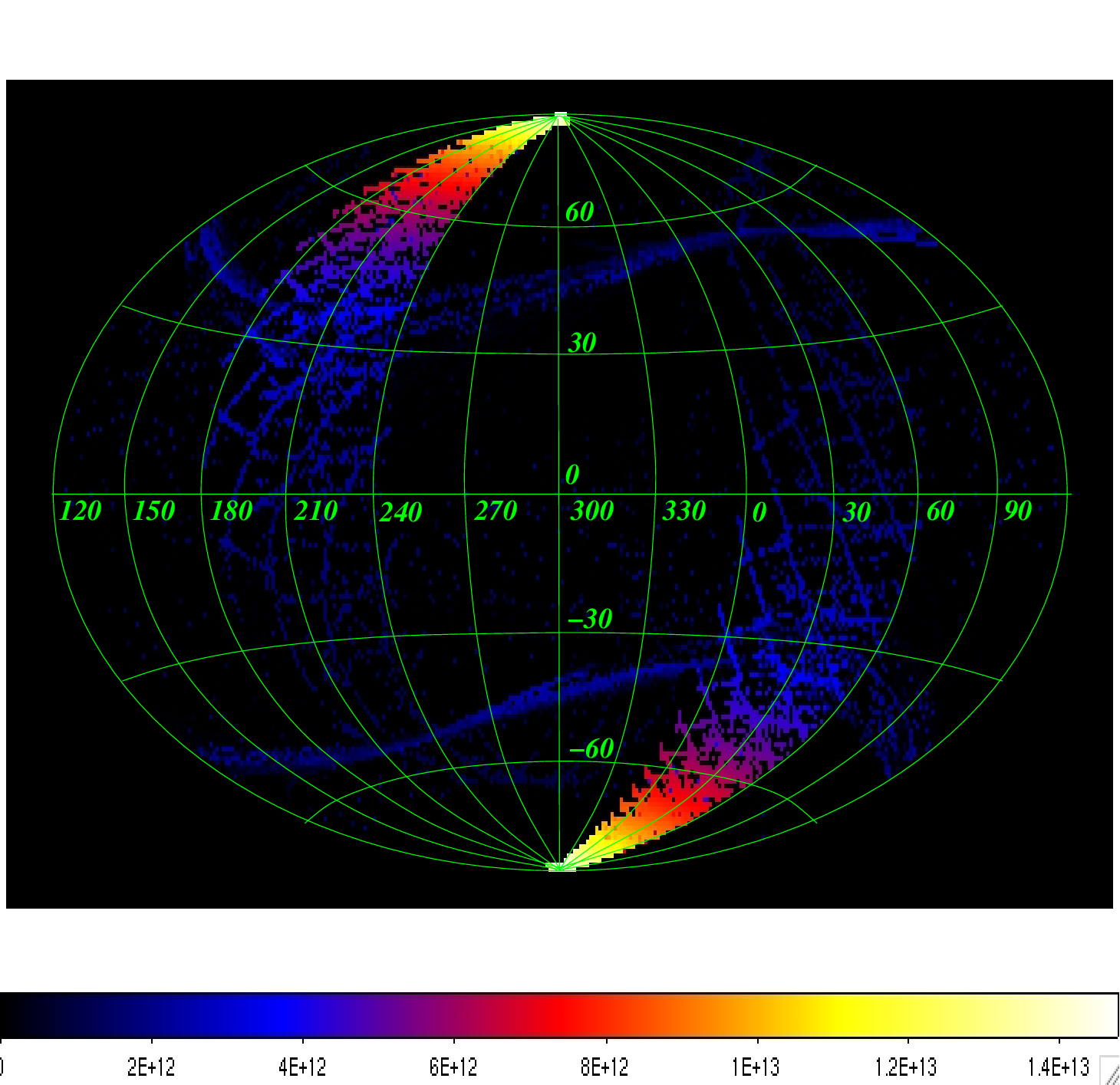}
\caption{Anisotropy pattern of the \gr\ emission from the gap.  The
  parameters are: $a=0.99GM$, $\chi=5^\circ$, $H=5R_{\rm
    Schw}$.Particles are initially injected in the equatorial plane at
  the distance $R=1.4r_H$ from the BH.}
\label{fig:map}
\end{center}
\end{figure}
 
The "direct" \gr\ emission from the vacuum gap is highly anisotropic
(in fact the opening angle of the emission cone is determined not only
by the width of the gap, but also by the deflections of the \gr\
trajectories by the BH gravitational field. Fig. \ref{fig:map} shows
the angular distribution of the average energies of the \gr s emitted
by particles accelerated in the gap. The highest energy photons are
emitted in the direction of the magnetic field. The in-spiralling
particles penetrate into the reconnection region mostly from the two
opposite directions in the equatorial plane, in which the reconnection
of the force-free surfaces starts at the largest distance (see the
bottom panels of Fig. \ref{fig:forcefree}). This explains the
appearance of the two emission regions at the opposite sides of the
black hole. The highest energy \gr s are emitted in the directions of
the southern and northern poles, along the direction of the magnetic
field. Lower energy photons emitted closer to the equatorial plane are
more strongly deviated and absorbed by the BH. A characteristic
anisotropy of the \gr\ emission from the gap is preserved even if the
magnetic field is not aligned with the BH rotation axis (see
e.g. \cite{neronov07}).

In most of the sources the magnetic field in
the gap is not aligned with the line of sight toward the
source, so that the direct \gr\ emission from the gap is not 
detectable. However, the 0.1-100~TeV \gr\ flux from the gap can be
efficiently "recycled" and "isotropized" via the development of
electromagnetic cascade in the infrared-to-ultraviolet photon
background in the source (see e.g. \cite{neronov07}). This should lead
to the redistribution of the primary \gr\ power over the lower energy
photons, from radio to softer \gr s. 

The details of the spectrum and anisotropy properties of the lower-energy electromagnetic emission from the cascades depend on the details of the soft photon and particle
distributions, as well as on the (largely uncertain) structure of magnetic fields in the cascade regions. Calculation of these details has to take into account observational constraints on the matter, radiation and magnetic field properties in the source and in its surrounding medium (the accretion flow, the AGN jet, the interstellar medium of the source host galaxy etc). Recent observations of the $\gamma$-ray emission from radio galaxies and blazars indicate that electromagnetic cascades initiated by the very-high-energy \gr s \cite{neronov02,neronov08a} and/or by the accelerated protons \cite{mannheim,bottcher,neronov08a} could play an important role in the physics of the AGN jets. 

\subsection{Power of direct \gr\ emission from the gap}

The UHECR data \cite{HiRes_GZK} imply that the total flux of the
cosmic rays above the cut-off at $E\sim 10^{20}$~eV is
about\footnote{This estimate is almost an order of magnitude lower
  than the one used in the Ref. \cite{neronov05} which relied on the
  AGASA data which did not show a signature of cut-off.}
\begin{equation}
F(E>10^{20}\mbox{ eV})\simeq (0.3\div 1)\times 10^{-12}\mbox{ erg/cm}^2\mbox{s}
\end{equation}
If this flux is produced by $N_{\rm source}\sim 100$ UHECR 
sources at the distances $D\sim 50\div 100$~kpc, the typical 
luminosity of an UHECR source is about
\begin{equation}
\label{source}
L_{\rm source}\simeq (10^{40}\div 10^{41})\left[\frac{100}{N_{\rm source}}\right]
\mbox{erg/s}
\end{equation}
Such an estimate should be compared to an estimate of the power of the
UHECR emission by a typical vacuum gap close to a supermassive BH in
AGN, which can be obtained in the following way.  At small inclination
angles $\chi$ the area of the region of the reconnection of the
force-free surfaces in the equatorial plane (see
Fig. \ref{fig:forcefree}) is estimated as ${\cal A}\sim 2\pi
r_H^2\sin\chi$. The maximal rate of injection of the charged particles
into the gap through the reconnection region is
\begin{equation}
R_{\rm max}\simeq n_{GJ}{\cal A}
\end{equation}
where $n_{GJ}$ is filled with the Goldreich-Julian density, given by
the Eq. \ref{GJ}. The energy of each particle ejected from the gap is
$E_{\rm curv}$, given by the Eq. \ref{curv_cut}. This means that the
maximal cosmic ray power of the gap is
\begin{equation}
\label{max}
L_{\rm CR,max}\simeq R_{\rm max}E_{\rm curv}\simeq 
6\times 10^{42}AZ^{-5/4}\left[\frac{M}{10^8M_\odot}\right]^{3/2}\left[\frac{B}{10^4\mbox{
G}}\right]^{5/4}\left[\frac{\chi}{1^\circ}\right]^{1/2}
\mbox{ erg/s}
\end{equation}
where we have assumed $\theta\simeq\chi$. Comparing the estimate
(\ref{source}) to (\ref{max}) one can see that the necessary cosmic
ray power of the vacuum gaps near the supermassive BHs can be achieved
already when the charged particles are injected into the gap at the
rate $\sim (0.01-0.1)R_{\max}$.

If the sources of UHECR are supermassive BHs in the centers of
galaxies, the acceleration to the energies above $10^{20}$~eV is
possible only in the ``loss saturated'' regime, in which the
acceleration rate is balanced by the energy loss rate, for most of the
trajectory of the particle. This means that the power of
electromagnetic emission from the gap dominates over the power emitted
in the form of the high-energy particles.

\begin{figure}
\begin{center}
\includegraphics[width=0.75\linewidth]{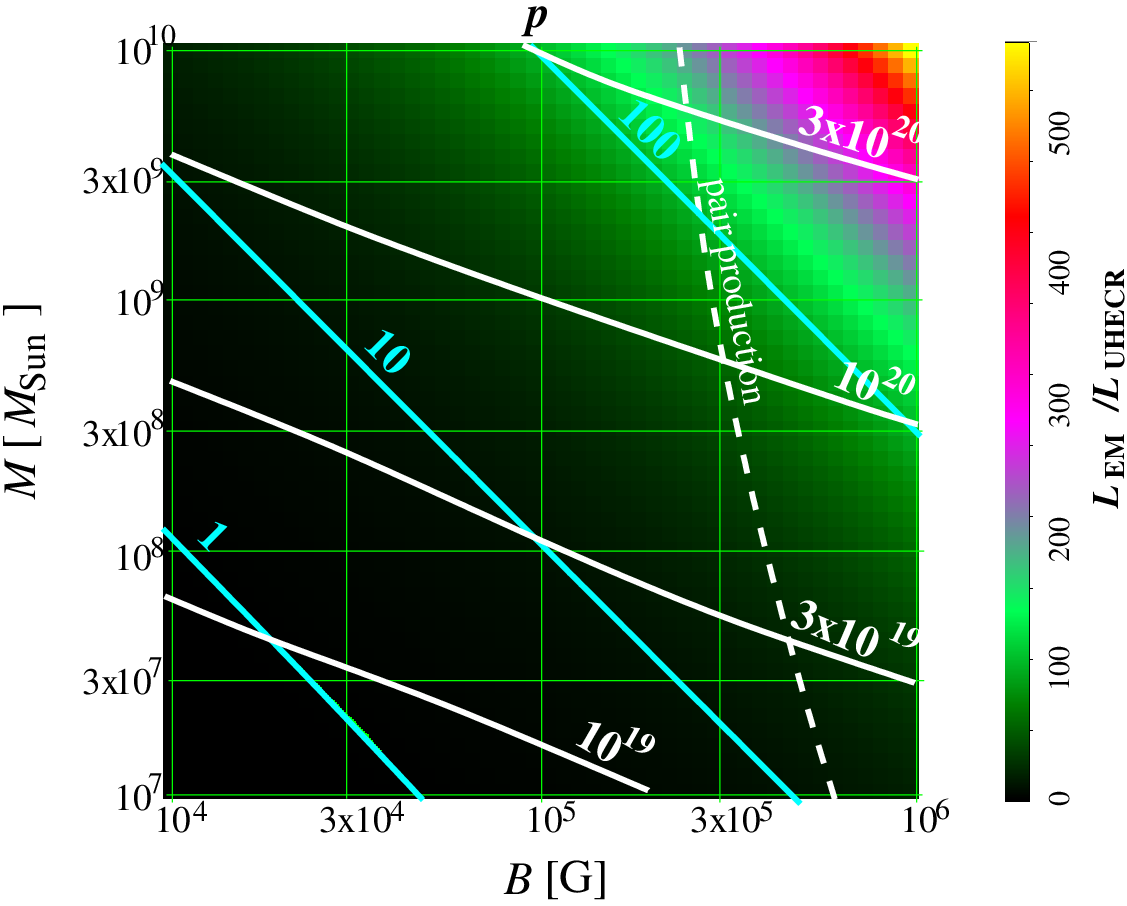}
\caption{Ratio of the electromagnetic to the cosmic ray power of the
  gap, as a function of the BH mass $M$ and magnetic field $B$. White
  contours show the maximal energies of the accelerated protons
  (see Fig. \ref{fig:E_vs_MB}). The parameters are:
  $a=0.99GM$, $\chi=5^\circ$, $H=5R_{\rm Schw}$. Particles are
  initially injected in the equatorial plane at the distance
  $R=1.4r_H$ from the BH. }
\label{fig:power}
\end{center}
\end{figure}

Even in the case of the almost aligned magnetic field, considered in
this paper, the power of electromagnetic emission from a
$(several)\times 10^{8}M_\odot$ BH is at least by a factor of $10\div
50$ larger than the UHECR power, as soon as the energy $10^{20}$~eV is
reached (see Fig. \ref{fig:power}). From this figure one can see that
the lines $L_{\rm EM}/L_{\rm UHECR}=const$ roughly correspond to
$M\sim B^{-3/2}$. This behaviour is readily explained, if one takes
into account that a simple estimate of $L_{\rm EM}/L_{\rm UHECR}$ can
be obtained by comparing the total available potential difference in
the gap to its fraction, spent on the particle acceleration (rather
than on the electromagnetic emission)
\begin{equation}
\frac{L_{\rm EM}}{L_{\rm UHECR}}\simeq \frac{E_{\rm max}}{E_{\rm cur}}\simeq 1Z^{5/4}A^{-1}\left[\frac{M}{10^8M_\odot}\right]^{1/2}\left[\frac{B}{10^4\mbox{ G}}\right]^{3/4}\left[\frac{\chi}{1^\circ}\right]^{1/2}
\label{EM/UHECR}
\end{equation}
(the above equation is valid if $E_{\rm cur}<E_{\rm max}$).

If the magnetic field is aligned with the rotation axis, $\chi\ll 1$,
the estimate of the electromagnetic power is much lower than the one
found for a general case of misaligned magnetic field in the
Ref. \cite{neronov05}. This means that a typical electromagnetic
luminosity of a nearby UHECR source is
\begin{equation}
L_{\rm EM,\ source}\ge (10\div 100)L_{\rm source}\sim 
(10^{41}\div 10^{43})\left[\frac{100}{N_{\rm source}}\right]
\mbox{erg/s}
\end{equation}
Such a luminosity is at the level of the luminosities of the AGN in
the local Universe.

The primary power emitted in the form of
the VHE \gr s is, most probably, recycled via the cascading on the
infrared backgrounds in the AGN central engine and in the host galaxy
of the source. The cascading leads to the isotropization of the
secondary emission, so that an observer is still able to detect a
certain fraction of the primary power in the form of lower energy
(e.g. synchrotron or inverse Compton) photons from the electromagnetic
cascade.

\subsection{Maximal height of the vacuum gap}
\label{sec:pair}

The VHE \gr s emitted by the accelerated particles can be absorbed in
interactions with soft background photons or with magnetic
fields. $e^+e^-$ pairs produced in these interactions could
initiate a development of electromagnetic cascade in the gap which
finally can ``short circuit'' the vacuum gap and
neutralize the parallel component of the electric field via
charge redistribution. Similarly to the case of vacuum gaps in the
pulsar magnetospheres, the condition that the high-energy gamma-ray
quanta emitted from the gap do not produce $e^+e^-$ pairs either in
interaction with the strong magnetic field or with the soft background
photons can limit the height of the gap. Requiring that the gap height
$H$ is smaller than the mean free path of a $\gamma$ ray through the
magnetic field, one finds \cite{erber}
\begin{equation}
\label{pair}
H\le D_{B\gamma}\approx
95\left[\frac{10^4\mbox{ G}}{B_\bot}\right] \exp(8m_e^3/3B_\bot\epsilon_{\rm curv})\mbox{ cm} 
\end{equation}

In the case when the magnetic field is almost aligned with the black
hole rotation axis, the \gr s are emitted along the direction of
propagation of the accelerated particles, which, in turn, move along
the magnetic field lines. For small inclination angles, the curvature radius of the magnetic field lines in the vacuum gap above the magnetic poles could be estimated as $R_B\sim R_{\rm Schw}/\sin(\chi)$. This means that the typical normal component
of the magnetic field encountered by the \gr s propagating inside the gap is
\begin{equation}
B_\bot\sim B\sin\chi
\end{equation} 
so that for the case $\chi\sim 1^\circ$, considered here, the magnetic
field component orthogonal to the particle and \gr\ velocity is some 2
orders of magnitude weaker than the parallel one. This means that even
if the magnetic field in the vacuum gap is very strong, $B\sim
10^6$~G, the typical mean free path of the \gr s is larger than the
characteristic distance scale of the problem (the Schwarzschild
radius, $R_{\rm Schw}$).  Thus, in the considered case of the gap with
``aligned'' magnetic field, the pair production is not important over
the entire range of BH masses studied in this paper. To check this
conclusion we have included in the numerical code a calculation of the
mean free path of the \gr s emitted at each step of the particle
trajectory and checked that the condition (\ref{pair}) is never
violated when the acceleration to the energies above $10^{20}$~eV is
possible in the range of parameters $M,B,H$ and $\chi$ considered
above.

Instead, in a realistic situation, the height of a vacuum gap with
aligned magnetic field is most probably limited by the large-scale
variations of the magnetic field, created by the accretion
flow. Assuming that the magnetic field has a certain inhomogeneity
scale $R_B$, one can find that the change of the field direction on
this distance scale will lead to the loss of alignment between the
trajectories of the \gr s emitted by particles accelerated in the gap
and the magnetic field, so that the normal component of the magnetic
field becomes $B_\bot\sim B$. We have calculated the magnetic field
strength at which the pair production on the distance scale $R_B$ can
set on, by substituting $B$ instead of $B\bot$ in the estimate of
Eq. \ref{pair} in numerical calculations. As a result, we have found
that the pair production can set on if the magnetic field strength is
above the threshold shown by a white dashed line in
Fig. \ref{fig:power}. One can see that in the case when the magnetic
field in the gap is aligned with the BH rotation axis, the pair
production on the distance scale of inhomogeneity of the magnetic fie
can set on at the magnetic field strengths, $B\ge 10^5$~G.

Electromagnetic cascade which limits the height of the gap can be initiated by the interactions of 
the \gr s not only with the magnetic field, but also with soft infrared photons emitted by the accretion flow. 
In order to estimate the importance of this  effect, one has to introduce additional assumptions about the radiative properties of the accretion flow. Curvature \gr s with the energies $E_\gamma\sim 1$~TeV produce $e^+e^-$ pairs mostly in interactions with the infrared photons of the energies $\epsilon\simeq 1\left[E_\gamma/1\mbox{ TeV}\right]^{-1}$~eV. Taking into account that the pair production cross section peaks at the value $\sigma_{\gamma\gamma}\simeq 10^{-25}$~cm$^{2}$, one can find that the optical depth of the gap for the TeV \gr s is
\begin{equation}
\label{taugammagamma}
\tau_{\gamma\gamma}=\frac{L_{\rm IR}\sigma_{\gamma\gamma}H}{4\pi R_{\rm IR}^2\epsilon}
\simeq 0.2\left[\frac{\left.L_{\rm IR}\right|_{\epsilon=1(E_\gamma/1{\rm\ TeV}){\rm\ eV}}}{10^{42}\mbox{ erg/s}}\right]
\left[\frac{R_{\rm IR}}{10^{16}\mbox{ cm}}\right]^{-2}\left[\frac{H}{10^{14}\mbox{ cm}}\right]\left[\frac{E_\gamma}{1\mbox{ TeV}}\right]
\end{equation}
Estimates of the optical depth of the gaps could be done if the constraints on the infrared luminosity of the AGN central engine and on the size of the infrared emission region are known from the observations, on the source-by-source basis (see e.g.  \cite{neronov07,neronov08a}). For example, from the above estimate one can see that a gap of the height $H\sim R_{\rm Schw}\sim 3\times 10^{13}$~cm in the magnetosphere of a $M\sim 10^8M_\odot$ BH surrounded by the accretion flow producing the infrared luminosity $L_{\rm IR}\sim 10^{42}$~erg/s in a region of the size $R_{\rm IR}\sim 100R_{\rm Schw}$ would be not closed in result of the $\gamma\gamma$ pair production inside the gap, since $\tau_{\gamma\gamma}<1$ in this case.

\section{Acceleration of heavy nuclei}
\label{sec:nuclei}

\begin{figure}
\begin{center}
\includegraphics[width=0.75\linewidth]{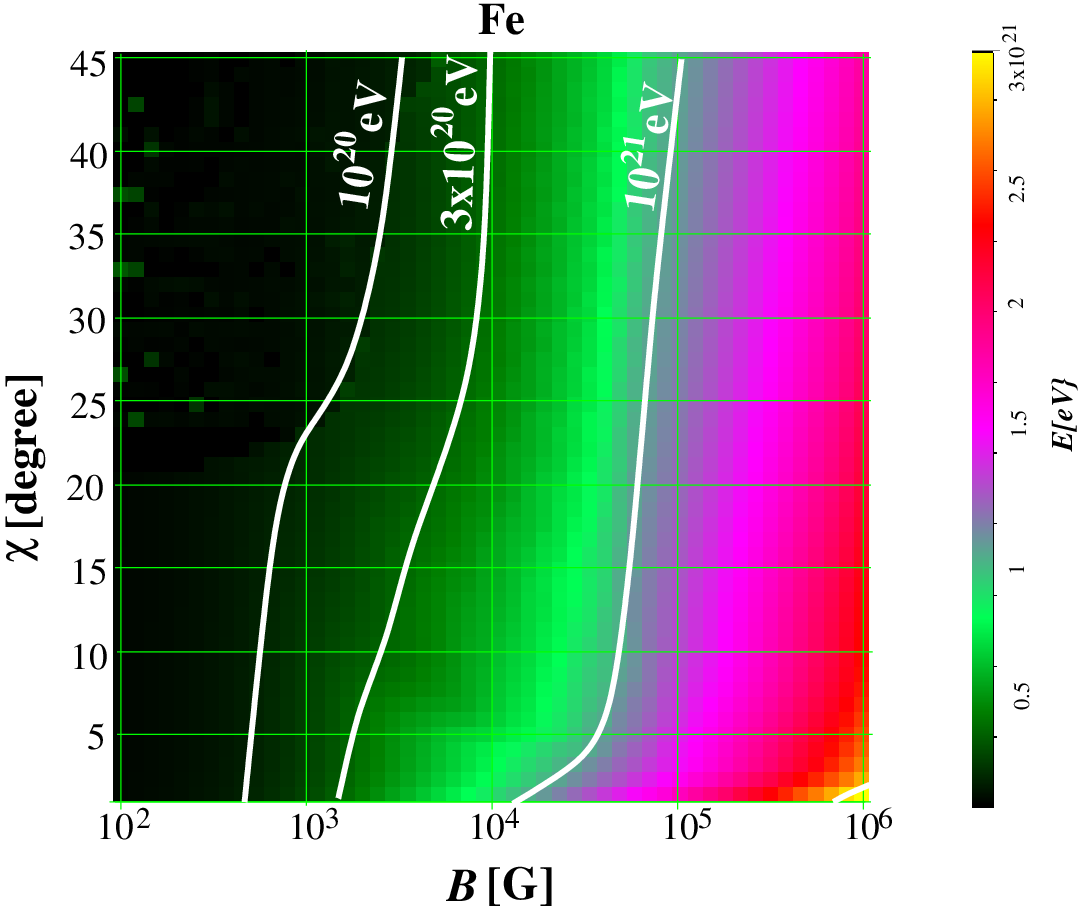}
\caption{Same as in Fig. 4 but for the case of acceleration of the iron nuclei. }
\label{fig:nuclei_E}
\end{center}
\end{figure}
\begin{figure}
\begin{center}
\includegraphics[width=0.75\linewidth]{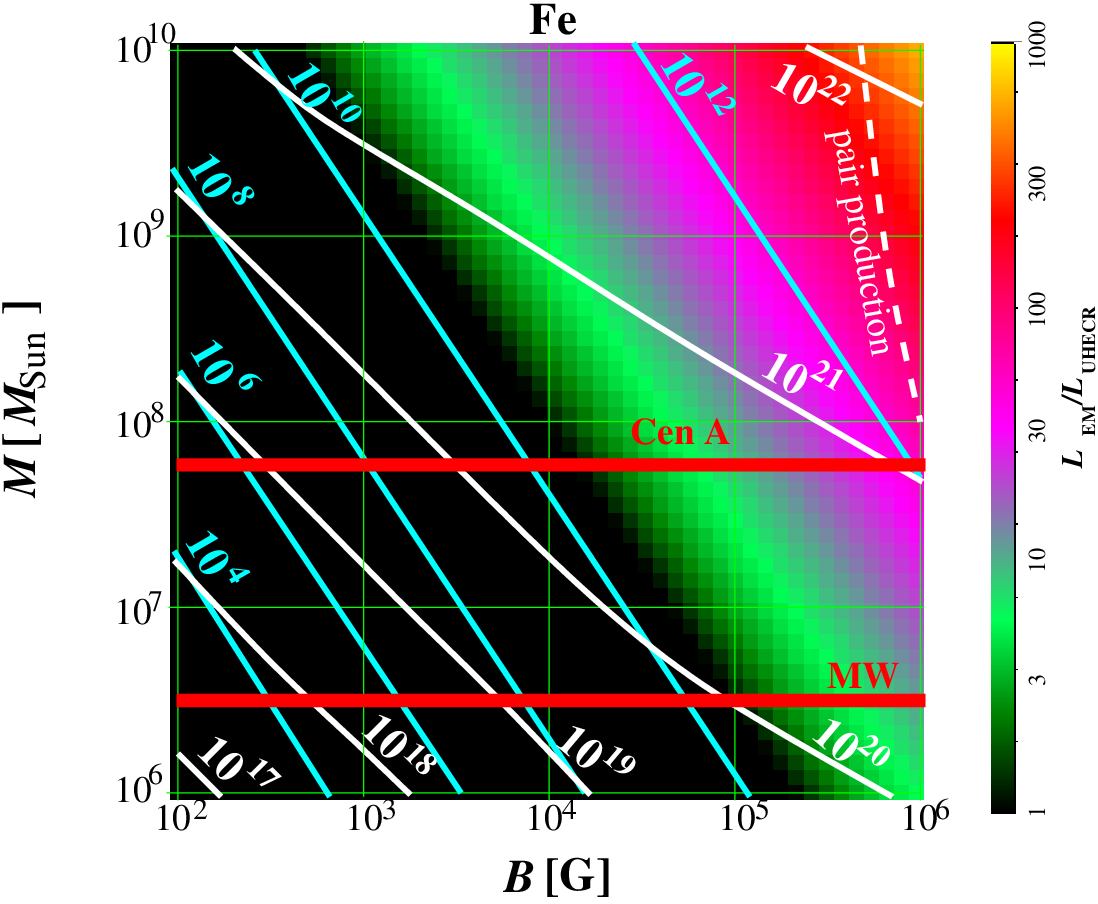}
\caption{Same as in Fig. \ref{fig:power} but for the case of
  acceleration of iron nuclei. The masses of the BH in the center of
  the Milky Way and in the nucleus of Cen A are marked by red
  horizontal lines. The white contours show the maximal energies of
  the accelerated Fe nuclei (in eV). Cyan contours show the maximal
  energies of the \gr s emitted by the nuclei (in eV). White dashed
  line shows the limit at which the pair production on the distance
  scale of inhomogeneity of the magnetic field sets on.}
\label{fig:nuclei_L}
\end{center}
\end{figure}

Up to now we have considered the acceleration of protons. However, our
results can be generalized, in a straightforward way, to the case of
acceleration of heavier nuclei, using the equations of Section
\ref{sec:maximal}.

Heavy nuclei, like the fully ionized iron, could be accelerated to higher
energies, since their charge is $Z>1$. However, nuclei with energies
above $10^{20}$~eV are easily disintegrated in interactions with the
infrared background photons which are present at the acceleration site,
 in the accelerator host
galaxy, in the intergalactic space and in the Galaxy. The secondary
nucleons produced in the photon disintegration, have the energy ${\it
  E}_N\simeq {\it E}/A$. Taking this into account, one can find that,
if the nuclei are fully disintegrated, the assumption about the
acceleration of nuclei does not result in the increase of the estimate
of the maximal energies of protons reaching the detector on the
Earth. E.g. the critical energy (per nucleon), at which the transition 
to the loss-dominated regime of acceleration happens (see Eq. \ref{ecrit}), 
\begin{equation}
E_{\rm crit,N}\simeq 
{\it E}_{crit}/A\sim A^{1/3}Z^{-2/3}
\end{equation} 
scales as $A^{1/3}Z^{-2/3}$, the factor which is less than 1 for most
of the primary nuclei. In the case of acceleration in the curvature
loss dominated regime, ${\it E}_{\rm cur,N}\simeq E_{\rm cur}/A\sim
Z^{-1/4}$ does not depend on the atomic number $A$ at all.

However, the assumption about full disintegration of the nuclei on the
way from the source to the observer on the Earth can be not valid in
some particular cases. First, if the energies of the Fe nuclei are
around $10^{20}$~eV, the mean free path of the nuclei through
the extragalactic infrared photon background is $\sim 100$~Mpc. This
means that the nuclei accelerated near the supermassive BH in the
nearby galaxies can cover the entire distance from the host galaxy to
the Milky Way without being photo-disintegrated
\cite{parizot}. Next, if the heavy nuclei are accelerated close to the
supermassive BH in the center of the Milky Way galaxy, they also can
reach the Earth without being photo-disintegrated. 

To verify if the heavy nuclei produced in a particular sources (supermassive black holes in the nearby AGN and normal galaxies) are likely to be photo desintegrated, one could notice that e.g. in the case of iron nuclei, the photo-desintegration cross-section, which peaks at a giant dipole resonance ($E_{\rm res}\sim 10-30$~MeV in the nucleus rest frame) is close to the $\gamma\gamma$ pair production cross-section, $\sigma_{\rm Fe-\gamma}\simeq 60(A-Z)Z/A$~mbar$\cdot$MeV$/E_{\rm res}\lesssim 10^{-25}$~cm$^2$ \cite{stecker99,khan04}. This means that the condition that the optical depth of the source with respect to the photo-desintegration is less than one imposes a constraint on the luminosity and/or size of the source, similar to the constraint following from the condition  $\tau_{\gamma\gamma}<1$ (see Eq. (\ref{taugammagamma})). Iron nuclei of the energy $E_{\rm Fe}\sim 10^{20}$~eV interact most efficiently with the far infrared photons of the energies $\epsilon_{\rm FIR}\simeq 3\times 10^{-2}\left[E_{\rm Fe}/10^{20}\mbox{ eV}\right]^{-1}$~eV, so that the restriction on the optical depth with respect to the photo-desintegration imposes a constraint on the size and luminosity of the source and of its host galaxy in the far infrared energy band.

The case of the acceleration of the heavy nuclei differs from the case
of proton acceleration in two important aspects: the higher mass of
the nucleus (and, respectively, lower Lorentz factor) result in the
reduction of the rate of the curvature energy loss and of the energies
of the quanta of curvature radiation. Because of the lower energies of
the curvature radiation quanta,the  pair production, which could limit the
size of the gap (see previous Section), sets on at higher values of the
magnetic field.  In addition, the electromagnetic luminosity of a
source accelerating nuclei to the energies $\sim 10^{20}$~eV could be
lower, than that of the source accelerating protons to $\sim
10^{20}$~eV.

Figs. \ref{fig:nuclei_E} and \ref{fig:nuclei_L} show the numerically
calculated maximal energies of accelerated particles and 
ratio of the electromagnetic to UHECR luminosity, as a
function of $M,B, \chi$, in the case of acceleration of Fe nuclei. The
region of parameter space in which the acceleration proceeds in the
``loss free'' regime (i.e. the estimate of the maximal energy is given
by Eq. \ref{potential}) is the black triangle in the lower left corner
of figure \ref{fig:nuclei_L}. One can see that in spite of the moderate BH masses
(down to $M\sim 10^7M_\odot$) and/or moderate magnetic fields (down to
$B\sim 10^3$~G in the case of a $10^9M_\odot$ BH), the acceleration up
to the energies $>10^{20}$~eV is possible in this regime.

As an example, let us consider the particularly interesting case of
particle acceleration near the supermassive BH in the center of the
Milky Way galaxy, which hosts a BH of the mass $M\sim 3\times
10^6M_\odot$ \cite{genzel} (horizontal red line in
Fig. \ref{fig:nuclei_L}). From Fig. \ref{fig:nuclei_L} one can see that
such a BH can accelerate iron nuclei up to the energies $\sim
10^{19}$~eV if the magnetic field around it reaches $B\sim 5 \times
10^3$ G. The energy loss in this case is much smaller then the power
of the cosmic ray emission $L_{\rm EM}/L_{\rm UHECR} = 5 \times
10^{-4}$. Fe nuclei with the energies $10^{18-19}$~eV produced in the
Galaxy are assumed to dominate the cosmic ray flux in this energy band
in certain models explaining the ``ankle'' feature of the cosmic ray
spectrum \cite{Fenuclei}. In order to explain the observed cosmic ray
flux at $10^{18}$ eV $F(10^{18}\mbox{ eV})\simeq 200$~eV/(cm$^2$s~sr)
\cite{HiRes_GZK}, the cosmic ray luminosity of a source in the
Galactic Center should be $L_{\rm Fe}\sim 3\times 10^{36}$~erg/s (the
luminosity can be still lower if the Fe nuclei diffuse through the
Galactic magnetic field). From Fig. \ref{fig:nuclei_L} one can see that
the power of electromagnetic emission, which accompanies iron nuclei
acceleration, could be as low as $L_{\rm EM}\le 10^{-3}L_{\rm Fe}$,
which is much less than the observed luminosity of the Galactic Center
in the TeV band ($\sim 10^{35}$~erg/s \cite{aharonian_GC}). Of course,
relaxing the assumption of approximate alignment of magnetic field
with the BH rotation axis would result in a higher estimate of the
electromagnetic luminosity. However, as it is clear from
Eq. \ref{EM/UHECR}, even in the loss-saturated regime of particle
propagation in the gap, the power of electromagnetic emission
increases by a factor of $\sim 10$ as the inclination angle of
magnetic field increases from several degrees to $\chi\sim 1$. Thus,
even in the absence of alignment of the magnetic field, the
electromagnetic emission which would accompany acceleration of the
iron nuclei to the energies $\sim 10^{18-19}$~eV close to the Galactic
Center BH would not be accompanied by significant electromagnetic
emission.

\section{Discussion and conclusions}
\label{sec:conclusions}

In this paper we have considered proton and heavy 
nuclei acceleration in the vicinity
of the horizon of the supermassive BHs. We have
shown that proton acceleration by a large scale electric field
induced by the rotation of the BH can result in
the production of cosmic rays with energies above $10^{20}$ eV,
provided that the magnetic field in the acceleration region is almost
aligned with the BH rotation axis (to better than few degrees, for a
reasonable range of black hole masses and magnetic field strengths
$10^8M_\odot\le M_{\rm BH}\le 10^{10}M_\odot,\ 10^4\mbox{ G}\le B\le
10^6\mbox{ G}$).

We have found that in a particular case of a rotating BH accreting
through an equatorial disk, the vacuum gaps, in which the large-scale
electric field is not neutralized by redistribution of charges, form
above the polar cap regions of the BH (see Fig.  \ref{fig:fields}). In
the case of magnetic field exactly aligned with the BH rotation axis,
charged particles from the equatorial accretion flow can not penetrate
into the polar cap regions (and, therefore, can not be accelerated).
However, as soon as the magnetic field is mis-aligned with the BH
rotation axis, the charged particles which are initially injected at
large distances into the equatorial plane, drift along the equatorial
plane toward the BH, until they reach the region of "reconnection" of
the force-free surfaces close to the BH horizon
(Fig. \ref{fig:forcefree}). As soon as the particles from the
equatorial plane penetrate into the reconnection region, they "leak"
into the vacuum gaps above the polar caps and are ejected to infinity
with high energies. Numerical modeling of particle transport and
acceleration close to the BH has enabled us to calculate the
dependence of the maximal attainable energies of the accelerated
particles, as well as of the properties of electromagnetic emission
produced by the accelerated particles on the physical parameters, such
as the BH mass $M$, the magnetic field $B$, the inclination of the
magnetic field $\chi$ and he gap height $H$ (see
Figs. \ref{fig:E_vs_H}, \ref{fig:E_vs_chi}, \ref{fig:E_vs_MB},
\ref{fig:E_vs_g}, \ref{fig:power}).

We have found that if the magnetic field is aligned to within several
degrees with the rotation axis, the electromagnetic luminosity of the gap is
just 10-50 times larger as compared its UHECR ($E>10^{20}$~eV) proton
luminosity.  A simple estimate has led us to a conclusion that the
resulting electromagnetic luminosity of the gap is of the order of the
bolometric luminosity of a typical AGN in the local Universe, which
makes the central engines of the AGN plausible candidates for the
astrophysical UHECR sources.

In the case of acceleration of heavy nuclei, acceleration to the
energy $\ge 10^{20}$~eV is possible for a much wider range of BH
masses and magnetic fields, without strong requirement on alignment
(see Fig. \ref{fig:nuclei_E}). In particular, even a BH of the mass
$\sim 3\times 10^6M_\odot$ (like the one in the center of the Milky
Way galaxy, \cite{BHMW}) can accelerate iron nuclei to the energies
above $10^{18}$~eV when the magnetic field is several hundred Gauss,
in the regime when the power of electromagnetic emission from the
accelerator is negligible, compared to its cosmic ray power.
Similarly, the nearest AGN Centaurus A, which hosts a BH with the mass
$\sim 6\times 10^7M_\odot$ \cite{BHCENA} can produce Fe nuclei with
the energies $10^{20}$~eV if the magnetic field close to the BH
horizon is above $3\times 10^3$~G. (see Fig. \ref{fig:nuclei_L}).  It is
important to understand how often the strong alignment between the
magnetic field and BH spin, $\zeta < (several)^\circ$, arises in
AGN. This will help to discriminate between proton and heavy nuclei at
highest energies, $E\ge 10^{20}$ eV.
  
The energy spectra of cosmic rays produced by individual sources via
the considered particle acceleration mechanism are sharply peaked at
nearly maximum energy, similarly to the spectra of high-energy particles
produced in linear accelerators. This feature can potentially
distinguish the presented mechanism from conventional shock
acceleration mechanism, which typically gives power law spectra
$dN/dE\sim E^{-\alpha}$ with $\alpha \ge 2$. The typical intrinsic
spectra of the UHECR sources can be found with a higher statistics of
the UHECR events, $N>100$ events at $E>60$ EeV
\cite{Kachelriess:2007bp}.

\end{document}